\documentstyle[12pt]{article}
\begin{document}
\def\F{{\cal F}}
\def\H{{\cal H}}
\def\E{{\cal E}}
\def\e{\epsilon}
\def\l{\lambda}
\def\O{\Omega}
\def\o{\omega}
\def\D{{\cal D}}
\def\d{\partial}
\def\t{\theta}
\def\T{\Theta}
\def\a{\alpha}
\def\ep{\varepsilon}
\def\P{\Pi_{p+1}}
\def\be{\begin{equation}}
\def\lb{\label}
\def\tdt{\tilde{\theta}}
\def\tdz{\tilde{z}}
\def\tdD{\tilde{{\cal D}}}
\def\tdd{\tilde{\partial}}
\def\G{{\cal G}}
\def\H{{\cal H}}
\def\bd{\bar{\d}}
\def\ee{\end{equation}}
\def\gg{{\bf g}}
\begin{titlepage}
\title{Paragrassmann Extensions of the Virasoro Algebra }
\author{\em  A.T.Filippov\thanks{Address until Jan.14, 1993: Yukawa
Institute of Theor. Phys., Kyoto Univ., Kyoto 606, Japan
{}~~~e-mail: filippov@jpnyitp~~~@jpnyitp.yukawa.kyoto-u.ac.jp
{}~~~filippov@theor.jinrc.dubna.su},
A.P. Isaev\thanks{e-mail address: isaevap@theor.jinrc.dubna.su} and
A.B.Kurdikov\thanks{e-mail address: kurdikov@theor.jinrc.dubna.su}\\ \\
\small\rm
Laboratory of Theoretical Physics,
JINR,  \\
\small\rm Head Post Office, P.O. Box 79  \\
\small\rm Dubna, SU-101  000 Moscow, Russia }
\date{}
\maketitle

\begin{abstract}

The paragrassmann differential calculus is briefly reviewed.
Algebras of the transformations of the para-superplane preserving
the form of the para-superderivative are constructed and their
geometric meaning is discussed. A new feature of these algebras is
that they contain generators of the automorphisms of
the paragrassmann algebra
(in addition to Ramond-Neveu-Schwarz - like conformal generators).
As a first step in analyzing these algebras we introduce more
tractable multilinear algebras not including the new generators.
In these algebras there exists a set of multilinear
identities based on the cyclic polycommutators.
Different possibilities of the closure are therefore admissible.
The central extensions of the algebras are given. Their number
varies from $1$ to $[\frac{p+1}{2}]$ depending on the form of
the closure chosen. Finally, simple explicit examples of the
paraconformal transformations are given.

\end{abstract}

\end{titlepage}

\newpage

\section{Introduction}
\setcounter{equation}0

Different extensions of the Virasoro algebra are useful in formulating
two-dimensional quantum conformal field theories with certain
additional symmetries [1].
Such extensions are generated by the stress-energy tensor $T$ and
some currents corresponding to the additional symmetries.
For example, if we add the Kac-Moody currents $J$ of the conformal
weight $1$, the result will be the well-known semi-direct sum of
the Virasoro and Kac-Moody algebras with Sugawara-type relations
between $J$ and $T$. Adding instead a fermionic current having
the weight $3/2$ we get the Ramond-Neveu-Schwarz algebra (RNS).
$SO(N)$ or $SU(N)$--invariant extensions and $N=2,3,4$
super-extensions of the RNS algebra have been considered
in the context of 2D conformal field theories in [2].

Introducing currents $W_{N}$ with integer weights $N \geq 3$
gives rise to the $W_{N}$-algebras of A.~Zamolodchikov which have
demonstrated their usefulness in last years.
So it seems that a most interesting new possibility is to look for
further extensions by trying to add currents
with fractional conformal weights and, in particular, parafermionic
currents of weights $(p+2)/(p+1)$ ($p$ is a positive integer
which is, in fact, the order of parastatistics or the degree of
nilpotency of an underlying paragrassmann
algebra). A construction of such an extension
of the Virasoro algebra is the subject of the present paper.

Attempts to realize this possibility had already been made. First of
all, we have to mention the work of V.~Fateev and A.~Zamolodchikov [3]
where a system with ${\bf Z}_{p+1}$--symmetry had been explored and
a certain associative operator `algebra of parafermionic currents'
has been constructed. That algebra, although, may hardly be
regarded as a paragrassmann extension of the Virasoro-RNS algebra
since it does not reproduce the RNS algebra for $p=1$~.

Another attempt, stimulated by a para-supersymmetric quantum mechanics
of V.~Rubakov and V.~Spiridonov [4], had been undertaken by S.~Durand
et.al.~[5]. The formulation of Ref.~[5] was based on a paragrassmann
calculus defined
in the frame of the Green representation for the paragrassmann
algebra also known as the Green {\it ansatz} [6]. In a later paper [7]
S.~Durand is using the paragrassmann calculus developed~\footnote{
As we have learned after submitting our paper [8] to the HEP-TH
database, some ingredients of this calculus for one variable were
earlier presented in Ref.~[9].} in Ref.~[8]
to derive interesting identities involving Virasoro-RNS - like
generators. Some of these identities look like plausible ingredients
of a para-extension of the superconformal algebra but their relation
to any symmetry transformations remains unclear. A relation of
similar identities to the conformal algebra in two dimensions, also
hinting at possible generalizations of the superconformal symmetry,
has been found by T.Nakanishi [10].

All this motivates our present attempt to find a systematic
para-generalization of the superconformal and Virasoro-RNS algebras.
To clarify the logic of our paper
we first recall the classical cases $p=0$ and $p=1$ in a framework
we are going to generalize.

$p=0$ ( {\it Virasoro case} )

The space is simply the complex plane with the coordinate $z$. Let
$F=F(z)$ be an analytic function of $z$, and consider an analytic map
\be
\lb{v1aa}
 z \mapsto \tilde{z} (z) \; , \;\;
F(z) \mapsto \tilde{F} \equiv F(\tilde{z} (z)) \; .
\ee
The trivial identity
\be
\lb{v3a}
\frac{\d}{\d z}\tilde{F} =  \tilde{z}' \frac{\d}{\d \tilde{z}}\tilde{F}
\ee
(prime will always denote the derivative with respect to$z$)
means that the old and new derivatives are proportional for any function
$\tilde{z}$, or in other words, any transformation (\ref{v1aa}) will be
conformal, i.e.  `preserving the form of the derivative'.
The group of the conformal transformations will be denoted by
${\bf CON}$.

Passing to the infinitesimal form of the transformation (\ref{v1aa}),
$ \tilde{z}=z+\l \o (z) \;$ ($\l$ is a small number),
and defining its generator $T(\o)$ by
\be
\lb{ni3}
\tilde{F} = (1+\l\;T(\o))\;F\;\;,
\ee
one easily sees that  $T(\o)=\o \d_{z} $.
The Lie algebra generated by $T(\o)$ is defined by the commutators
\be
\lb{v5}
[T(\o),T(\psi)]=T(\o \psi^{\prime}-\o^{\prime}\psi)\;
\ee
and is called the conformal algebra, $Con$ in our notation.
It coincides with the whole algebra of vector fields on the complex
plane and its (unique) central extension is the standard Virasoro
algebra denoted by $Vir$.

There is a simple generalization of the construction when $F$ are
not ordinary functions but conformal fields of weight
$\Delta$. In this case the transformation rule is
\be
\lb{v4a}
F_{\Delta}(z) \mapsto \tilde{F}_{\Delta} \equiv
( \tilde{z}' )^{\Delta} F_{\Delta}(\tdz)\;\;,
\ee
and the generators have the form  $T(\o)=\o \d_{z} + \Delta \o' $.
Their Lie algebra coincides with (1.4).

$p=1$ ( {\it RNS-case} ).

Now the space is a complex superplane with coordinates $z$ and $\t$,
$\t^{2}=0$. In fact, for a precise formulation we need a Grassmann
algebra of more than one variable.
One of them will be specified as $\t$ while the rest will be
referred as `other thetas'. A super-analytic map is
\begin{eqnarray}
 z  \mapsto  \tdz & = & Z_{0}+\t Z_{1}\;, \lb{ni11} \\
\t  \mapsto  \tdt & = & \T_{0}+\t \T_{1}\;,\;\;\tdt^{2}=0\;,
 \lb{ni12}
\end{eqnarray}
where $Z_{i}$ and $\T_{i}$ are functions of $z$ and of `other
thetas' with needed Grassmann parities.
This map transforms functions $F=F(z,\t)$ into
 $ \tilde{F}\equiv F(\tilde{z},\tilde{\t})$.

The fractional derivative of order $1/2$ can be defined as
\be
\lb{v8}
{\D}=\d_{\t}+\t\d_{z}\;\;, \;\; {\D}^{2}=\d_{z} \;\;,
\ee
due to the Grassmann relations $\t^{2}=0=\d^{2}_{\t}\;,\;
\{\d_{\t},\t \}_{+} =1$.
The super-analytic maps (\ref{ni11}),
(\ref{ni12}) are called superconformal transformations
when the superderivative transforms homogeneously (see e.g. [11], [12]).
This requirement, similar to (\ref{v3a}), can be written in the form
\be
\lb{v9}
{\D}\tilde{F}= \Phi \tilde{{\D}}\tilde{F}\;\;,
\ee
and leads to certain restrictions on the parameter functions, namely,
$$
 \Phi = \D \tilde{\t} \; ,
\D \tilde{z} = \D ( \tilde{\t}) \tilde{\t} \; , \; \;
$$
or, in the component notation,
\be
\begin{array}{lll}
Z_{1}  & = & \T_{1} \T_{0}\;, \\
Z_{0}' & = & \T_{0}'\T_{0} + (\T_{1})^{2}\;.
 \end{array}
 \lb{ni14}
 \ee
These transformations form a group which is called superconformal,
or ${\bf CON}_{1}$ in our notation. This group is well-studied in
connection with the theory of the superconformal manifolds and is the
main tool in constructing two-dimensional superconformal field theories
[11], [12]. One can solve the equations (\ref{ni14}) and so find
finite superconformal transformations. However, the calculation uses
anticommutation relations between $\t$ and other thetas and it is not
easy to generalize to arbitrary $p$. Anyway, it is much easier to work
with infinitesimal transformations even in the Grassmann case.

For the infinitesimal form of the mapping (\ref{ni11}), (\ref{ni12}),
\be
\lb{v6}
\begin{array}{lll}
 \tilde{z} & = & z +\; \l \O(z,  \t ) =
 z+ \;\l (\o_{0} +\t \o_{1}  )\;\;,\;\;  \\
\tilde{\t} & = & \t + \;\l \E (z, \t )
 = \t+ \;\l (\e_{0} +\t \e_{1} )\;\;,
\end{array}
\ee
the conditions (\ref{ni14}) read
\be
\lb{ni15}
\o_{1}=\e_{0}\;\;,\;\;\o_{0}'=2\e_{1}\;.
\ee
Thus, defining the generators ${\cal T}(\o_{0})$ and $\G(\e_{0})$ by
\be
\lb{ni16}
\tilde{F} = (1+\;\l({\cal T}(\o_{0})+\G(\e_{0})))\;F\;\;,
\ee
one can easily find that
\be
\lb{v11}
{\cal T}(\o)=\o\d_{z}+\frac{1}{2}\o^{\prime}\t\d_{\t}\;\;,\;\;
\G(\e)=\e (\d_{\t} - \t\d_{z})\;\;.
\ee
These generators close into the well-known Lie algebra
\be
\lb{v12}
\begin{array}{lll}
\left[ {\cal T}(\o),{\cal T}(\psi)\right] & = & {\cal T}(\o
\psi^{\prime}-\o^{\prime}\psi)\;,	 \\
\left[ {\cal T}(\o),\G(\e) \right] & = & \G(\o
\e^{\prime}-\frac{1}{2}\o^{\prime}\e)\;,  \\
\left[ \G(\e),\G(\zeta)\right] & = & {\cal T}(\e \;\zeta ) \;.
\end{array}
\ee
that we denote by  ${\cal CON}_{1}$.

As in the previous example, it is possible to introduce similar
generators (\ref{v11}) for the superconformal field $F$ of arbitrary
weight $\Delta$:
$F(z,\t) \mapsto (\D \tdt )^{\Delta} F(\tilde{z}, \tilde{\t} )$
(cf. with (\ref{v4a})). The corresponding generators,
\be
\lb{v11a}
{\cal T}(\o)=\o\d_{z}+\frac{1}{2}\o^{\prime}\t\d_{\t} +
\frac{\Delta}{2} \o' \;\;,\;\;
\G(\e)=\e (\d_{\t} - \t\d_{z}) - \Delta \e' \t \;\;,
\ee
obey the same algebra (1.15).
The operator $Q=(\d_{\t} - \t \d_{z} )$ which appeared in the
definition of $\G(\e)$ anticommutes with $\D$ and is called
the supersymmetry generator.

Up to now, $\o$ and $\e$ were respectively even and odd functions
of the variables $z$ and of `other thetas', i.e. some polynomials
in `other thetas' with coefficients being analytic functions of $z$.
However, all $\theta$ variables anticommute\footnote{Note that the
commutation relations between $\omega$, $\epsilon$, and $\theta$
are completely fixed by Eq.~(1.9) and by the condition
$\tilde{\theta}^2=0$.} and so we can move all parameters containing
`other thetas' to left-hand sides of all equations.
This obviously allows to introduce `bare' generators $T$ and $G$
whose arguments $\omega$ and $\epsilon$ are ordinary functions of $z$
independent of `other thetas'. Then the expressions for $T$ and $G$
coincide with (1.9) and obey the algebra (1.15) with the last
commutator replaced by anticommutator. This algebra (or, more precisely,
the superalgebra) is what is usually called
the `superconformal algebra'. We will denote it by $Con_{1}$ and
its unique central extension, the Ramond-Neveu-Schwarz algebra,
is denoted by $Vir_{1}$.

The distinction between the algebras ${\cal CON}_{1}$ and $Con_{1}$
is usually ignored, being practically trivial.
As we shall see in Sect.3, this is not the case for their para-analogs
${\cal CON}_{p}$ and $Con_{p}$
since the paragrassmann algebra of `other thetas' is a much more
complicated object than the Grassmann one (and, probably, not
uniquely defined, see [13]).
In particular, we do not know at the moment simple and general
commutation rules between the elements of the paragrassmann algebra.
As a result, closing the algebra
${\cal CON}_{p}$ becomes a rather non-trivial problem.
On the contrary, the algebra  $Con_{p}$
closes quite easily, even in several non-equivalent variants, if the
number $p+1$ is rich in divisors.
Each variant of closing defines an extended algebra, $Vir_{p}$,
with a number of central charges (from 1 to $[(p+1)/2]$).

The simplest variant of $Vir_{p}$ looks as follows:
\begin{eqnarray}
\;[L_{n}\;,\;L_{m}] & = & (n-m)L_{n+m}+
\frac{2}{p+1}\left(\sum_{j}c_{j}\right)
		      (n^{3}-n)\delta_{n+m,0}\;,\nonumber   \\
\;[L_{n}\;,\;G_{r}] & = & (\frac{n}{p+1}-r)G_{n+r}\;,\lb{v1a}   \\
\;\{G_{r_{0}},\dots ,G_{r_{p}}\}_{c} & = & (p+1)L_{\Sigma r}-
      \sum_{j}c_{j}\left(\sum_{i}r_{i}r_{i+j}+\frac{1}{p+1}\right)
			\delta_{\Sigma r,0}\;,\nonumber     \\
j & = & 1\dots \left[\frac{p+1}{2}\right]\;,\nonumber
\end{eqnarray}
where $L_{n}=T(z^{-n+1})\;\;,\;\;G_{r}=G(z^{-r+\;1/(p+1)})$, and
$\{ \dots \}_{c}$ is the cyclic sum of the $(p+1)$-linear monomials:
$$
\{ G_{0} \; , \dots \; , G_{p} \}_{c} =
G_{0} \cdots G_{p} + G_{p}\cdot G_{0} \cdots G_{p-1} +
\dots +G_{1} \cdots G_{p}\cdot G_{0} \; .
$$
Note that a particular variant of this algebra $Vir_{p}$,
with totally symmetric bracket in the third line and
without central extensions, had been presented in Ref.[14]
under the name `fractional Virasoro algebra'.
Recently, a generalization that relates the algebras of
Refs. [5] and [14], has been given in Ref.[7].

The rest of the paper is devoted to a generalization of the previous
scheme to arbitrary integer $p$.
In Sect.2, a necessary preliminary technique of the paragrassmann
algebras is summarized and its versions most useful for constructing
paraconformal transformations are presented. A detailed description of
the differential calculus with one and many variables is given in
Ref.[13].

In Sect.3, we first discuss main properties of the fractional derivative
${\D}$ (${\D}^{p+1}=\d_{z}$) which are valid in any version of the
paragrassmann calculus.
Then, in the spirit of the above scheme, we introduce paraconformal
transformations to construct a paraconformal group ${\bf CON}_{p}$
and corresponding algebras ${\cal CON}_{p}$, and $Con_{p}$ (Sect.4).
We show that a $p$-analog of the infinitesimal transformations
(\ref{v6}) must look as
\be
\lb{nip1}
\begin{array}{rcl}
\tilde{\t} & = & \t + \;\l \E (z, \t )\;,    \cr
 \tilde{z} & = & z +\; \l \O^{(1)}+\dots + \l^{p} \O^{(p)}\;.
\end{array}
\ee
Unlike the Lie algebras (and superalgebras) having only first order
generators, here we have to retain `higher-order generators'
thus introducing  into consideration  a $p$-jet structure.
This suggests that the algebra  ${\cal CON}_{p}$
might be a $p$-filtered Lie algebra containing the generators
of $p$ `generations', $\{{\cal L}^{(i)}\}$,
so that an analog of the formula (\ref{ni16}) would look like
$$
\tilde{F} = (1+\lambda \{ {\cal L}^{(1)} \} +
\lambda^{2}\{ {\cal L}^{(2)} \} + \dots +
\lambda^{p}\{ {\cal L}^{(p)} \})\;F\;\;.
$$
The algebra  ${\cal CON}_{p}$ contains generators of a new type
(we call them ${\cal H}$-generators) that do not act on $z$ but are
crucial in closing the algebra. The algebras $Con_{p}$ can be closed
without them.

In Sect.5, we briefly discuss  the  meaning of the construction
in terms of algebraic geometry.
This allows us to introduce central charges in a straightforward
way and so to derive the algebras $Vir_{p}$.
A discussion of the properties of these algebras, of possible
generalizations, and of unsolved problems is given in Sect.6.

Appendix presents an explicit formulas for some paraconformal
transformations for $p=2$ generalizing corresponding superconformal
transformations. These formulas make more clear the analogies and
differences between geometry of the superplane and that of the
para-superplane.

Concluding this rather long introduction we would like to point out
that the main formal results of this paper were known to us for some
time~\footnote{A preliminary version of this paper appeared under the
same title as the preprint JINR E5-92-393, Dubna, 1992. The present text
is somewhat rearranged, made more concise, and new results have been
added to Sect.~3 and Appendix.} and have been presented at seminars and
workshops this spring. However, we refrained from publishing them prior
to understanding their geometric meaning. We hope that we can now
suggest a possible geometric foundation for our formal construction
in terms of the versions (`version covariance') and of jet-like
structures although much remains to be done to completely uncover
a geometric meaning of paraconformal transformations.

\section{Paragrassmann Algebra $\P$ }
\setcounter{equation}0

In Ref.~[8] we have considered paragrassmann algebras
$\Gamma_{p+1}(N)$ with $N$ nilpotent variables $\theta_{n}$,
$\theta_{n}^{p+1}=0$, $n=1,\ldots,N$. Some wider algebras
$\Pi_{p+1}(N)$ generated by $\theta_{n}$ and additional
nilpotent generators $\d_{n}$ have also been constructed.
These additional generators served for defining a paragrassmann
differentiation and paragrassmann calculus. The building block
for this construction was the simplest algebra $\Pi_{p+1}(1)$.
By applying a generalized Leibniz rule for differentiations in
the paragrassmann algebra $\Gamma_{p+1}(N)$
we have found two distinct realizations for $\Pi_{p+1}(1)$ closely
related to the $q$-deformed oscillators.
We have mentioned in [8] that other realizations of the $\Pi_{p+1}(1)$
may be constructed. A complete list of these realizations
({\it versions}) and a fairly general approach to constructing algebras
$\Pi_{p+1}(N)$ have been presented in Ref.[13]. Here we reproduce the
results of this work that are essential for understanding the main body
of the present paper. Especially important is the fact that, under
certain conditions, all these realizations are equivalent and one
may choose those which are most convenient for particular problems.

The algebra $\P(1)$ is generated by the nilpotent variables $\t$ and
$\d$ satisfying the conditions
\be
    \lb{u3}
\t^{p+1} = 0  =  \d^{p+1} \;\;,
\ee
(it is implied, of course, that $\t^{p}\not=0$ and the same for $\d$).
Any version of the algebra $\P(1)$ is defined by the relation allowing
to move $\d$ to the right of $\t$

\be
\lb{u7}
\d \t = b_{0} + b_{1}\t \d +b_{2}\t^{2} \d^{2} + \dots +
b_{p}\t^{p} \d^{p} \; ,
\ee
where $b_i$ are complex numbers restricted by consistency of
the conditions (\ref{u3}) and (\ref{u7}) and by further assumptions
to be formulated below. With the aid of Eq.~(\ref{u7}) any element
of the algebra can be expressed in terms of the basis
$\theta^r \d^s$, i.e. in the normal-ordered form.
This relation obviously preserves the natural grading in the associative
algebra generated by $\theta$ and $\d$ satisfying (\ref{u3})
\be
\lb{u4}
deg\;(\t^{r_{1}}\d^{s_{1}}\t^{r_{2}}\d^{s_{2}}\dots
 \t^{r_{k}}\d^{s_{k}}\;) = \Sigma r_{i} - \Sigma s_{i}\;,
 \ee

A useful alternative set of parameters, $\a_{k}$, also fixing
the algebra may be defined
\be
\lb{u8}
\d \t^{k} = \a_{k}\t^{k-1} + (\dots )\d\; ,
\ee
where dots denote a polynomial in $\theta$ and $\d$. This relation is
a generalization of the commutation relation for the standard
derivative operator, $\d_z z^k = k z^k +z^{(k-1)} \d_z$, and we may
define the differentiation of powers of $\theta$ by analogy,
\be
\lb{pa14i}
\d (\t^{k}) = \a_{k}\t^{k-1} , \; \alpha_0 \equiv 0 \; ,
\ee
to be justified later.

By applying Eq.~(\ref{u7}) to Eq.~(\ref{u8}) one may derive the
recurrent relations connecting these two sets of the parameters:
\be
\lb{pa14} \\
\a_{k+1}  =  \sum^{k}_{i=0} b_{i} \frac{(\a_{k})!}{(\a_{k-i})!} \; ,
\ee
where $(\a_{k})! \equiv  \a_{1}\a_{2} \cdots \a_{k}$.
These relations enable us to express $\a_{k}$ as a function of
the numbers $b_{i}\;,\; 0 \leq i \leq k-1$ .
The inverse  operation, deriving $b_{i}$ in terms of $\a_{k}$,
is well-defined only if all $\a_{k}\not=0$.

The consistency condition mentioned above is that the parameters
must be chosen so as to satisfy the identity
 $$0 \equiv \d \theta^{p+1}\;.$$
Taking into account that the second term in Eq.~(\ref{u8})
vanishes for $k= p+1$ we have $\a_{p+1}= 0$, with no other restrictions
on the parameters $\alpha_k$ with $k \leq p$. The corresponding
restriction on $p+1$ parameters $b_i$ follows from Eq.~(\ref{pa14}),
\be
\lb{u9}
\a_{p+1}(b_{0},\dots ,b_{p})\equiv
 b_{0} + b_{1}\a_{p} + b_{2}\a_{p}\a_{p-1} + \dots +
b_{p}\a_{p}\a_{p-1} \cdots \a_{2}\a_{1} =0 \; ,
\ee
where the parameters $\alpha_i$ are expressed in terms of $b_i$.
Any admissible set $\{b\}$ determines an algebra $\P^{\{b\} }$
with the defining relations (\ref{u3}), (\ref{u7}).
To each algebra $\P^{\{b\} }$ there corresponds a set $\{\a\}$.
{\it A priori}, there are no restrictions on $\{\a\}$,
but, if we wish to treat $\d$ as a non-degenerate derivative
with respect to $\t$, it is reasonable to require,
in addition to (\ref{u9}), that
\be
\lb{u10}
all \;\;\a_{k} \not= 0\;.
\ee
So let us call a set $\{b\}$ (and corresponding algebra$\P^{\{b\} }$)
{\it non-degenerate}, if the condition (\ref{u10}) is fulfilled,
and degenerate otherwise.
As it was already mentioned, in the non-degenerate case the numbers
$b_{i}$ are completely determined by the numbers $\a_{k}$,
so we can use the symbol $\{\a\}$ as well as $\{b\}$.
An interesting fact is that nontrivial algebras (${\a}_1 \neq 0$)
may be degenerate if and only if $p+1$ is a composite number [13].
We will see later that the arithmetic properties of $p+1$ are also
important in  constructing paraconformal transformations.

In general, different sets $\{b\}$ determine non-equivalent algebras
$\P^{\{b\} }$ and, at first sight,
the algebras corresponding to different sets $\{b\}$  look very
dissimilar. However, this is not true for the non-degenerate ones.
In fact, all non-degenerate algebras $\P^{\{b\} }$ are isomorphic to
the associative algebra $Mat(p+1)$ of the complex $(p+1) \times (p+1)$
matrices.

The isomorphism can be manifested by constructing an explicit
exact (`fundamental') representation for $\P^{\{b\}}$ . With this
aim we treat $\t$ and $\d$ as creation and annihilation operators
(in general, not Hermitian conjugate) and introduce the ladder of
$p+1$ states $|k\rangle $, $k=0,1, \ldots ,p$ defined by
\be
\lb{pa8}
\d |0\rangle = 0 \;, \; |k\rangle \sim \t^{k} |0\rangle \; , \;
\t |k\rangle = \beta_{k+1} |k+1\rangle \; .
\ee
Here $\beta$'s are some non-zero numbers, reflecting the freedom of
the basis choice. As $|p+1\rangle =0$, the linear shell of
the vectors $|k\rangle$ is finite-dimensional
and in the nondegenerate case, when all $\beta_{k} \neq 0 \;\;\;
(k=1, \dots ,p)$, its dimension is $p+1$.

Using (\ref{pa8}) and (\ref{u8}) we find
\be
\lb{pa11}
\d |k\rangle = (\a_{k} /\beta_{k}) |k-1\rangle \;\; .
\ee
Thus the fundamental (Fock-space) representations of the operators
$\t$ and $\d$ is
\begin{eqnarray}
\lb{pa13a}
\t_{mn} & = & \langle m| \t |n\rangle =\beta_{n+1} \delta_{m,n+1} \;
\; , \\
\d_{mn} & = & \langle m| \d |n\rangle =(\a_{n} / \beta_{n} )
\delta_{m,n-1} \; \; .
\lb{pa13b}
\end{eqnarray}
It is not hard to see that for non-zero $\a$'s,
the matrices corresponding to $\t^{m}\d^{n}\;\;(m,n=0\dots p)$,
form a complete basis of the algebra $Mat(p+1)$. The isomorphism
is established.

Thus, different non-degenerate algebras  $\P^{\{b\} }$
are nothing more than alternative ways of writing
one and the same algebra $\P$.  We are calling them {\it versions}
having in mind that fixing the $b$-parameters is somewhat analogous to
a gauge-fixing (by adopting a broad meaning of this term introduced by
H.~Weyl in his famous book on quantum mechanics).

This implies that we will mainly be interested in `version-covariant'
results, i.e. independent of a version choice.  Nevertheless, special
versions may have certain nice individual features making them
more convenient for concrete calculations (thus allowing for simpler
derivations of covariant results by non-covariant methods).

The same is true about the matrix representations. In principle, we
need not use any matrix representation for the paragrassmann variables
and derivatives. However, in some calculations the existence of the
exact matrix representation (\ref{pa13a}), (\ref{pa13b}) is very useful
for deriving version-covariant identities in the algebra $\P$. For
instance, using the representation it is easy to check that
\begin{eqnarray}
\{ \d\;,\;\t^{(p)}\} & = &
\left(\sum \a_{k}\right)\;\t^{p-1}\;,       \lb{u12}
\cr
\{ \d^{p}\;,\;\t^{(p)}\} & = & \prod \a_{k}\;,      \lb{u13}
\end{eqnarray}
and to find many other relations. Here we have introduced the notation
\be
\lb{u14}
\{ \Xi \; ,\;\; \Psi^{(l)} \} = \Xi \Psi^{l} + \Psi \Xi \Psi^{l-1}
+ \dots + \Psi^{l} \Xi \;
\ee
that will be often used below.
The identities (\ref{u12}) generalize those known in the
para-supersymmetric quantum mechanics [4].

Note also that one may adjust the parameters  $\beta_{k}$
to get a convenient matrix representation for $\t$ and $\d$.
As a rule, we take $\beta_{k} =1$. For the versions with real
parameters $\a_{k}$, it is possible to
choose $\beta_{k}$ so as to have $\t^{\dag}=\d$ .
We also normalize $\t$ and $\d$ so that $\a_{1}\equiv b_{0}=1$.

Now consider three special versions that are most suitable for
constructing paraconformal transformations and correspond to simplest
forms of Eq.~(\ref{u7}).

(1): {\bf q-Version or Fractional Version} \\
Here $ b_{1}=q \neq 0 \; , \;\; b_{2}=b_{3}= \dots =b_{p}=0 \; , \;
{\rm so \;\; that}$
$$
\a_{i}=1+q+ \dots +q^{i-1} = \frac{1-q^{i}}{1-q} \; .
$$
The condition $\alpha_{p+1}=0$ tells that $q^{p+1}=1 \; \; (q \neq 1)$,
while the assumption that all $\a_{i} \neq 0$ forces $q=b_1$ to be
a primitive root, i.e. $q^{n+1} \neq 1 \; , \;\; n<p$ . Thus, in this
version ($\d =\d_{(1)}$)
\be
\d_{(1)} \t = 1+ q\t \d_{(1)} \; ,
\lb{pa16}
\ee
$$
\d_{(1)}(\t^{n})=(n)_{q} \t^{n-1} \; ,
\;\; (n)_{q}=\frac{1-q^{n}}{1-q} \; .
$$
These relations were derived in Ref.~[8] by assuming that
$\d$ is a generalized differentiation operator, i.e.
satisfying a generalized Leibniz rule (a further generalization is
introduced below). The derivative $\d_{(1)}$ is naturally related to
the $q$-oscillators ($q$-derivative) and to quantum algebras;
Eq.~(\ref{pa16}) is also most convenient for
generalizing to Paragrassmann algebras with many $\t$ and $\d$
(see [8], [13] and references therein).

(2): {\bf Almost Bosonic Version} \\
For this Version
$$
 b_{1}=1 \; , \;\; b_{2}= \dots =b_{p-1}=0 \; , \;\;
b_{p} \neq 0 \; , \;\; {\rm so \;\; that} \;\; \a_{k}=k
$$
The condition $\alpha_{p+1}=0$ gives $b_{p}=-\frac{p+1}{p!} \;$and thus
\be
(\d_{(2)})_{mn}=n\;\delta_{m,n-1} \; , \;\;
\d_{(2)} \t = 1+ \t \d_{(2)} -\frac{p+1}{p!}\; \t^{p} \d_{(2)}^{p}\;.
\lb{pa17}
\ee
This derivative is `almost bosonic' as
$\d_{(2)}(\t^{n}) = n \t^{n-1} \;\; (n \neq p+1 ) \; $. This version
is convenient for rewriting the generators of para-extensions of the
Virasoro algebra in the most concise form in the next section.

Let us now discuss the interrelations between $\t$ and $\d$.
As we have already mentioned the notation itself hints at treating
$\d$ as a derivative with respect to $\t$ (see (\ref{u8}).
To be more precise, let us represent an arbitrary vector
$|F \rangle = \sum_{k=0}^{p} f_{k} | k \rangle $
as the function of $\theta$
$$
F (\t) = \sum_{k=0}^{p} f_{k} \t^{k}.
$$
The action of the derivative $\d$ on $F(\t)$
is defined by (\ref{pa8}) and (\ref{pa11}) ($\beta_{k} =1$),
\be
\lb{pa21a}
\d (1)=0 \;,\;\;\d (\t^{n}) = \a_{n}\t^{n-1}\;\;(1 \leq n \leq p)\;.
\ee
It is clear, however, that this derivative does not
obey the standard Leibniz rule $\d(ab)=\d(a)b + a\d(b)$.

So consider the following  modification of the Leibniz rule [8],
[15]
\be
\lb{pa22}
\d(FG) = \d(F) \bar{g} ( G) + g (F) \d(G) \; .
\ee
The associativity condition (for differentiating $FGH$) tells that
$g$ and $\bar{g}$ are homomorphisms, i.e.
\be
\lb{pa23}
g(FG)=g(F)g(G) \; , \;\; \bar{g}(FG)=\bar{g}(F)\bar{g}(G) \; .
\ee
The simplest natural homomorphisms compatible with the relations
(\ref{pa21a}), (\ref{pa22}), and (\ref{pa23}) are linear
automorphisms of the algebra  $\Gamma_{p+1}$,
\be
\lb{pa24}
g(\t)=\gamma \t \; , \;\; \bar{g}(\t)=\bar{\gamma} \t \; ,
\ee
where $\gamma \; , \; \bar{\gamma}$ are arbitrary complex parameters and
\be
\lb{pa25}
\a_{k} = \frac{{\bar{\gamma}}^{k}-\gamma^{k}}{\bar{\gamma}-\gamma}.
\ee
Using the condition (\ref{u9})
and assuming nondegeneracy, $\a_{n} \neq 0 \; (n < p+1)$,
we conclude that $\bar{\gamma}/\gamma $ must be a primitive $(p+1)$-root
of unity. Thus we may formulate another interesting version
of the paragrassmann algebra $\P$

(3):{\bf $g-\bar{g}$--Version} \\
In this version the parameters $\alpha_k$ are given by Eq.~(\ref{pa25}),
and we can calculate $b_i$ by solving Eq.~(\ref{pa14}):
$b_{0}=1, \; b_{1} = \bar{\gamma} + \gamma -1, \;
b_{2} = (\bar{\gamma} -\bar{\gamma}\gamma + \gamma -1)/
(\bar{\gamma} + \gamma), \; \dots $. \\
Here $\gamma$ and $\bar{\gamma}$ are complex numbers constrained by the
condition that  $q=\bar{\gamma}/\gamma$ is a primitive root of unity.
 From Eqs.~(\ref{pa22}) and (\ref{pa24}) one can derive the following
operator relations for the automorphisms $g, \; \bar{g}$
\be
\lb{pa27}
\d \t -\gamma \t \d = \bar{g}\;,\;\;\d \t -\bar{\gamma}\t \d = g\;.
\ee

For the special case $\gamma = (\bar{\gamma})^{-1} = q^{1/2}$
identifying $\d = a, \; \t = a^{\dag}$
allows to recognize in (\ref{pa27}) the definitions of the
$q$-deformed oscillators in the Biedenharn-MacFarlane form [16].
Version-(1) can be derived from Version-(3) by
putting $\bar{\gamma}=q, \; \gamma=1$ (or $\bar{\gamma}=1, \;\gamma=q$).
For $p=2$ all the above versions satisfy the modified Leibniz rule
(\ref{pa22}).
However, for $p > 2$ Version-(2) satisfies a different modification of
the Leibniz rule
\be
\lb{pa30}
\d(FG) = \d(F) \bar{\gg} (G) + \gg (F) \d(G) + Lz(F,G) \; .
\ee
For Version-(2) $\bar{\gg} = \gg = 1$, and the additional term
$Lz(.\,,.)$ belongs to the one dimensional space $\{ | p \rangle \}$.
We suggest to call this term  the `Leibnizean'.

This modification as well as other versions are
discussed in our accompanying paper [13]. There we also have constructed
paragrassmann calculus with many variables.
Note that a reasonably simple many-variable paragrassmann calculus
with a generalized Leibniz rule can be formulated only for Versions
(1) and (3). Nevertheless, Version-(2) is also useful
as will be demonstrated below.

Before we proceed it is important to realize the following.
While the Grassmann calculus satisfying our requirements is
unique\footnote{For example, commuting Grassmann variables used in the
Green {\it ansatz}
are forbidden by the requirement that any linear combination of the
Grassmann variables must also be a Grassmann variable, i.e. nilpotent.},
it is probably not true for many paragrassmann variables. We have proved
that all nondegenerate algebras $\P(1)$ are equivalent but we have no
corresponding theorem for the algebras $P(N)$ constructed in Ref.[13],
and
it is also possible that our construction does not exhaust many-variable
algebras. For these reasons, in what follows we try to avoid, as much as
possible, using any particular many-variable calculus. Then at some
points we have to do some guesswork instead of precise computations and
this  results in some gaps in our geometric reasoning.
Being fully aware of this,
we still think that the geometric approach is indispensable for
understanding paraconformal algebras, and so present it in its
incomplete form. We hope that the gaps can be filled later as far as the
many-variable paragrassmann calculus will be fully developed. Note also
that the multilinear algebra $Vir_p$ though not precisely derived from
the algebra $Con_p$ has a self-dependent interest.  With  these
cautionary remarks, we turn to our main task.

\section{Paraconformal Transformations}
\setcounter{equation}0

To start realizing the program outlined in Introduction consider
a para-superplane ${\bf z}= (z , \; \t )$, where
$z \in { \bf C}$ and $ \t$ is the generator of the paragrassmann
algebra $\Gamma_{p+1}(1)=\Gamma_{\t}$, i.e. $\t^{p+1}=0$.
Any function defined on this plane has the form
\be
\lb{c1}
F \equiv F(z, \; \t) = F_{0}(z) + \t F_{1}(z) +
\t^{2}F_{2}(z) + \dots + \t^{p}F_{p}(z).
\ee
It is useful to define an analog of the superderivative as a
$(p+1)$-root of the derivative $\d_{z}$ [8], [9]
\be
\lb{c2}
{\D} = \d_{\t} +\kappa \frac{\t^{p}}{(\a_{p})!} \d_{z} \; , \;\;\;
\D^{p+1} = \kappa\d_{z} \; .
\ee
We denote here the $\theta$-derivative in arbitrary version by $\d_{\t}$
instead of $\d$ and shall often use this notation to make some formulas
more transparent. The number $\kappa$ will be fixed later.

The action of this operator on the function (\ref{c1}) is
\be
\lb{c3}
\D F(z, \; \t) = F_{1}(z) + \a_{2}\t F_{2}(z) +  \dots +
\a_{p}\t^{p-1}F_{p}(z) + \kappa \frac{\t^{p}}{(\a_{p})!}F_{0}'(z) \; ,
\ee
where $F' = \d_{z} F$. In analogy with the super-calculus  [11], [12]
 the inverse operator $\D^{-1}$ (defined up to a constant independent of
$z$ and $\t$)
$$
\D^{-1} F \equiv
\int^{\bf z}  \; d{\bf z} \; F =
\frac{(\a_{p})!}{\kappa}\int^{z} \; dz' \; F_{p}(z') + \t F_{0}(z) +
\frac{\t^{2}}{\a_{2}}F_{1}(z) +  \dots +
\frac{\t^{p}}{\a_{p}}F_{p-1}(z) \;
$$
may be formally interpreted as an `indefinite' integral.

Now we can formally introduce a `definite'
integral~\footnote{The integral
over $z$ is to be understood as a contour integral in the complex
$z$-plane, and it is contour-independent as far as the contours are
not crossing singularities of $F_{p}(z)$. Thus the integral might be
regarded as a `contour' integral in a para-superplane. If there is
no singularity inside a closed contour, the integral must be zero.}
\be
\lb{c3aa}
\int^{{\bf z}_{1}}_{{\bf z}_{2}} \; d{\bf z} \; F =
\int^{{\bf z}_{1}}  \; d{\bf z} \; F -
\int^{{\bf z}_{2}}  \; d{\bf z} \; F \; .
\ee
Of course, to make this definition and formulas below completely
meaningful
we have to use a well-defined paragrassmann calculus for many
variables~\footnote{In fact we need an embedding of the algebra
$\Gamma_{p+1}(1)$ into some infinite dimensional
paragrassmann algebra $\Gamma_{p+1}(1+ \infty )$ with
generators $\t_{0}=\t, \; \t_{1}, \t_{2}, \dots \; $ .}.
We instead are using some natural formal rules which have to be valid
in different versions of the many-variable calculus. We assume that the
action of the derivative $\d_{\t}$ on the expression having $\t$ on the
extreme left can be calculated by Eq.(2.4) and that the derivative of
any expression containing no $\t$-variable is zero (this is a nontrivial
assumption as one can find considering examples of many-variable
calculus [8], [13]). We will also as long as possible avoid assumptions
on commuting different paragrassmann variables.

Now we can construct an analog of the Taylor expansion.
Successively using the relation
$$
F({\bf z}_1) - F({\bf z}_2) =
\int^{{\bf z}_{1}}_{{\bf z}_{2}} \; d{\bf z} \; \D F({\bf z})
$$
and the identity
$$
{\D}^n F({\bf z}) = {\D}^n F({\bf z}_2) + [{\D}^n F({\bf z}) -
{\D}^n F({\bf z}_2)]
$$
one can write a generalized Taylor expansion in the form
\be
\lb{tay1}
F({\bf z}_1) = \sum_{n}I_{n}({\bf z}_1 ,
{\bf z}_2) {\D}^n F({\bf z}_2)\; ,
\ee
where $I_0 =1$ and
\be
\lb{tay2}
I_{n+1}({\bf z}_1 , {\bf z}_2) =
\int^{{\bf z}_{1}}_{{\bf z}_{2}} \; d{\bf z} \;
I_{n}({\bf z} , {\bf z}_2) \; .
\ee
The first $p$ integrals are easy to compute:
$$
I_1 = \t_{1}-\t_{2} \; ,
$$
$$
I_2 = \frac{1}{\a_{2}}\t_{1}^{2}-\t_{1}\t_{2} +
(1-\frac{1}{\a_{2}})\t_{2}^{2} \;\; , \;\; \dots
$$
$$
I_{p+1} =
\frac{1}{\kappa}(z_{1}-z_{2}) -\frac{1}{(\a_{p})!}\t_{1}^{p}\t_{2} +
(1-\frac{1}{\a_{2}})\frac{1}{(\a_{p-1})!}\t_{1}^{p-1}\t_{2}^{2} +
\dots \; .
$$

In view of Eq.(\ref{c2}) and by analogy with the Grassmann case
[11], [12] one might expect that all the higher integrals can be
expressed in terms of the first $p+1$ integrals and that these
integrals are translation invariant.
This is not quite the case for arbitrary $p$, the generalization is not
so naive and requires some care. For not to digress from our main goal
we postpone a detailed treatment of the integrals to a separate
publication.
The simplest case $p=2$ is briefly outlined in Appendix after
introducing a generalization of super-translations, which is also a
nontrivial problem.

We think that these properties of $\D$ and $\D^{-1}$ justify regarding
${\D}$ as a correct generalization of the superderivative and we
proceed with realizing our program. Thus consider invertible
transformations of the para-superplane
\be
\begin{array}{rcl}
 z & \rightarrow & \tdz(z,\t)  \; , \;\;\; deg(\tdz)=0 \; ,  \\
 \t & \rightarrow & \tdt(z,\t)  \; , \;\;\; deg(\tdt)=1 \; ;
\end{array}
\lb{glob}
\ee
\begin{eqnarray}
\tdz & = & Z_{0}(z) + \t Z_{1}(z) + \dots +\t^{p} Z_{p}(z) \; ,
\lb{tdz}\\
\tdt & = & \T_{0}(z) + \t \T_{1}(z) + \dots + \t^{p} \T_{p}(z) \; ,
\lb{tdt}
\end{eqnarray}
where $deg$ is a natural ${\bf Z}_{p+1}$-grading in
$\Gamma_{p+1}(1+\infty )$ and $Z_{i}$ and $\T_{i}$ are  functions
of $z$ with values in $\Gamma_{p+1}(1+\infty)/\Gamma_{\t}$
of needed grading. Here we assume that it is possible
to move all $\t$ to the left-hand sides of the
para-superfields (\ref{tdz}) and (\ref{tdt}) (for Version-(1) it is
evident). We also require that
\be
\lb{c5} \tdt^{p+1}=0
\ee
and, of course, $[\tilde{z} , \tilde{\t}] = 0$.

The corresponding transformation for the functions (\ref{c1}) is
defined as
\be
\lb{c6}
\tilde{F} \equiv \tilde{F}(z,\t) = F(\tdz,\tdt)
= F_{0}(\tdz) + \tdt F_{1}(\tdz) + \dots + \tdt^{p}F_{p}(\tdz) .
\ee
In accord with (\ref{c3}), $\tilde{\D}\tilde{F}$ is
\be
\lb{c7}
\tilde{\D}\tilde{F} =  F_{1}(\tdz) + \tdt F_{2}(\tdz) +
\dots + \frac{\kappa}{(\a_{p})!}\tdt^{p}F_{0}'(\tdz).
\ee

Following the route described in Section~1,  consider the
transformations obeying the requirement analogous to Eq.~(\ref{v9})
\be
\D F(\tdt, \tdz) = \Phi(\t, z)\tdD F(\tdt, \tdz ) \;\;,
\ee
or, in the operator form,
\be
\lb{DD}
\D = \Phi\tdD .
\ee
Acting on $\tdt$ we immediately get
\be
\lb{dd}
\Phi = \D \tdt \;\;.
\ee

Consider main properties of these transformations. They satisfy the
group property since for two sequential transformations
${\bf z} \mapsto  \tilde{\bf z}({\bf z})
\mapsto \tilde{\tilde{\bf z}}(\tilde{\bf z}({\bf z}))$
we have (see (\ref{DD}), (\ref{dd}))
\be
\lb{DDD}
\D = (\D\tdt)\tdD = (\D\tdt)(\tdD\tilde{\tdt})\tilde{\tdD}
= (\D\tilde{\tdt})\tilde{\tdD} .
\ee
This group will be called `paraconformal' and referred to as
${\bf CON}_{p}$.

 The condition (\ref{DD}) is a very strong restriction on possible form
of transformation functions (\ref{glob}).
One can show that all the restrictions can be derived by putting
$F$ to be $\tdt^{k}$, $\tdz$ and $\tdt\tdz$. This gives
\begin{eqnarray}
\lb{Dtdtk}
\D\tdt^{k} & = & \a_{k}(\D\tdt)\tdt^{k-1}\;,  \\
\lb{Dtdz}
\D\tdz & = & \frac{\kappa}{(\a_{p})!} (\D\tdt)\tdt^{p}\;, \\
\lb{Dtdtz}
\D(\tdt\tdz) & = & (\D\tdt)\tdz\;.
\end{eqnarray}
Note that (\ref{Dtdz}) allows us to interpret
(\ref{DD}) as the rule for differentiation of composite functions
\be
\lb{Ddd}
\D = (\D\tdt)(
\tilde{\d_{\t}} + \kappa\frac{\tdt^{p}}{(\a_{p})!}\tilde{\d_{z}} ) =
(\D\tdt)\tilde{\d_{\t}} + (\D\tdz)\tilde{\d_{z}} .
\ee
Summarizing the main properties of the derivative $\D$ and of the
transformations satisfying
 (\ref{DD}), we conclude that it is reasonable to regard them as
 the paraconformal transformations of the para-superplane.

The main restrictions on the parameters of paraconformal transformations
are coming from Eq.~(\ref{Dtdz})~\footnote{Eqs.(\ref{Dtdtk}),
(\ref{Dtdtz})
give additional restriction, mostly on the commutation properties of the
parameters.} that, taken in components,
give rise to the following relations between $Z_{i}$ and $\T_{j}$
(defined by (\ref{tdz}), (\ref{tdt}))
\be
\lb{kurd1}
\begin{array}{rcl} Z_{1} & = &
 \frac{\kappa}{(\a_{p})!}\T_{1}\T^{p}_{0} \; , \\
\t Z_{2} & = & \frac{\kappa}{(\a_{p})!} ( \t \T_{2}\T^{p}_{0} +
 \frac{1}{\a_{2}} \T_{1} \{ \t\T_{1}\;\;,\;\; \T^{(p-1)}_{0} \} ) \;,\\
	& \dots & \\ \t^{p-1} Z_{p} & = & \frac{\kappa}{(\a_{p})!}
( \t^{p-1} \T_{p}\T^{p}_{0} + \dots + \frac{1}{\a_{p}}
\T_{1} \{ (\t\T_{1})^{(p-1)}\; , \;\; \T_{0} \} ) \; , \\
\t^{p} Z^{\;\prime}_{0} & = &
\frac{\kappa}{(\a_{p})!} \t^{p} \T_{0}'\T^{p}_{0} + \a_{p}\t^{p-1}
\T_{p}\{ \t\T_{1}\;\; , \;\;
\T^{(p-1)}_{0} \} + \dots + \T_{1}\cdot (\t \T_{1})^{p} \; .
\end{array}
\ee
Here the notation introduced in (\ref{u14}) is used.

These relations generalize the much  simpler relations (\ref{ni14})
to any $p$. It is rather hard to push forward the analysis with
complicated expressions (\ref{kurd1}) without having a well-established
technique for handling many thetas. For this reason, we are forced to
turn to the infinitesimal language. Then, introducing a small $c$-number
parameter $\lambda$ one may rewrite the transformations
(\ref{glob} -- \ref{tdt}) as
\begin{eqnarray}
 \tdz(z,\t) & = & z + \lambda \O(z,\t)+ O(\lambda^{2}) \;, \;\;
\O  = \sum^{p}_{i=0} \t^{i} \o_{i}(z)  \; ,
   \lb{c4} \\
 \tdt(z,\t) & = & \t + \lambda {\cal E} (z,\t) + O(\lambda^{2}) \;,
\;\;\E  =  \sum^{p}_{i=0} \t^{i} \e_{i}(z) \; .
   \lb{c4a}
\end{eqnarray}
where $\o_{i}$ and $\e_{i}$ are first coefficients in expansions of the
functions $Z_{i}(z)$ and $\T_{i}(z)$ in powers of $\lambda$.
{\it A priori}, there is no reason to exclude the higher
powers from consideration.
The infinitesimal transformations (\ref{c4}), (\ref{c4a})
satisfying the paraconformal conditions (\ref{Dtdtk} -- \ref{Dtdtz})
define a linear space which we denote by  ${\cal CON}_{p} \;$.
This is an infinitesimal object corresponding to the
paraconformal group ${\bf CON}_{p}$ and so it must carry some
algebraic structure induced by the group structure of ${\bf CON}_{p}$.
In this sense, we will speak of `the algebra   ${\cal CON}_{p}$'
though its algebraic properties will be discussed later.
Here we briefly analyze a geometric meaning of  ${\cal CON}_{p}$.

As it is evident from (\ref{c4a}), the only component of $\tdt$
containing a finite part is  $\T_{1} = 1 + \lambda \e_{1}(z)$;
the rest of $\T_{i} $ are  of the first order in $\lambda$.
Then (\ref{kurd1}) tells that only $Z_{0}$ and $Z_{p}$
contain the terms of the first order in $\lambda$, while the other
$Z_{i}$ with $1 \leq i \leq p-1 $ must be of the order
$\lambda^{p+1-i}$. This suggests that all terms up to the order
$\lambda^{p}$ must be kept in (\ref{c4}), (\ref{c4a}).
So, since the transformed function is generally defined as
$
\tilde{F} = (1 +
 (general\;\; element\;\; of\;\; the\;\; algebra))F\;,
$
(cf. (\ref{ni16}))
a general element of the algebra  ${\cal CON}_{p}$  must be of the form
\be
\lb{genel}
(general \;\; element) = \lambda \{ {\cal L}^{(1)} \} +
\lambda^{2}\{ {\cal L}^{(2)} \}+ \dots +
\lambda^{p}\{ {\cal L}^{(p)} \}\;.
\ee
Here we denote by  $\{ {\cal L}^{(M)} \}$
a set of the generators of the $M$-th {\it generation}.
Eq.(\ref{kurd1}) shows that $M$-th generation  $\{ {\cal L}^{(M)} \}$
must contain some new generators
$\{ {\cal X}^{(M)} \}$ that are not present in $\{ {\cal L}^{(M-1)} \}$.
This would guarantee, in particular, the appearance of non-zero
$Z_{p+1-M}$.
If it were possible to put  all $Z_{i}\;\;(i=1\dots p-1)$
to zero we would get rid of these subtleties. But as we will see below,
this  contradicts to the requirement of the bilinear closure of the
algebra ${\cal CON}_{p} \;$. In fact, new generators naturally arise
 from certain bilinear brackets of the old ones,
and this process stops only at the $p$-th generation.

 Returning to geometry, the formula (\ref{genel}) indicates
that the algebra ${\cal CON}_{p}$ lives not in the tangent space of the
group but rather in a space of $p$-jets (see, e.g. Ref.[17]).
For the $p$-jets the generators must be of the form
$(something)\d^{j}\;\;(j=1\dots p)$. This is right the case for the
algebra ${\cal CON}_{p}$, as we will see
below\footnote{In geometric terms,
one has to consider tangent bundles of the order $p$.
There exist a natural
relation between $\Gamma_{p+1}(1)$ and tangent bundles of the order $p$
studied in the context of A.Weil's bundles of infinitely near points
(see, e.g. Ref.[18]). Although this relation might be useful also
in our case, we apparently need to deal with more complex structures.}.

Now an explicit realization of the paraconformal
algebra generators ${{\cal L}}^{(M)}$ is in order.
We concentrate on the first generation
because generators of the higher generations
must be obtained through multilinear combinations of them.
Substituting (\ref{c4}) (\ref{c4a})
in (\ref{Dtdtk}) -- (\ref{Dtdtz}) we get the (first-order smallness)
infinitesimal form of the paraconformal conditions:
\begin{eqnarray}
\D \{\E, \t^{(k)}\} & = & (k+1)((\D \E)\t^{k}+ \{\E, \t^{(k-1)}\})
\;,\;\;k=1 \dots p-1    \;,                    \lb{DEk}    \\
\D \O & = & \frac{\kappa}{(\a_{p})!}
((\D \E)\t^{p} + \{\E, \t^{(p-1)}\})\;,        \lb{DO} \\
\D (\E z + \t \O ) & = & (\D \E) z + \O\;\;.   \lb{DOE}
\end{eqnarray}

Eqs.(\ref{DEk} -- \ref{DOE}) lead to certain restrictions on the
functions $\e_{i}, \;\; \o_{i} $
(we assume  that $\o_{0}$ and $\e_{1}$ are ordinary functions,
free of any paragrassmann content). The condition (\ref{DO}) gives
\be
\lb{o123}
\o_{1} = \o_{2} = \dots =\o_{p-1} = 0 \; ,
\ee
 \be
  \e_{1} = \frac{1}{p+1} \o_{0}' \; ,
\ee
\be
\a_{p}\t^{p-1}\o_{p} =\frac{\kappa}{(\a_{p})!} \{\e_{0},\t^{(p-1)}\} \;,
\ee
wherefrom, by virtue of the relation $\{\e_{0} \; , \;\; \t^{(p)}\} = 0$
that follows from the nilpotency condition (\ref{c5}), we have
\be
\lb{oep}
\t^{p}\o_{p} = -\frac{1}{\a_{p}}\frac{\kappa}{(\a_{p})!} \e_{0}\t^{p}\;.
\ee
 From the third relation (\ref{DOE}) we find that      %and (\ref{o123})
\be
 \lb{ope}
  \t^{p}\o_{p} = \frac{\kappa}{(\a_{p})!} \t^{p}\e_{0}\;,
   \;\;{\rm or } \;\; \o_{p} = \frac{\kappa}{(\a_{p})!} \e_{0}\;,
 \ee
 which, together with (\ref{oep}), gives the commutation relation
\be
\lb{c21}
\e_{0} \t^{p} + \a_{p} \t^{p} \e_{0} = 0 \;.
\ee
The condition (\ref{DEk}) provides the commutation rules of $\e_{i}$
with $\t^{k}$ and $\d$:
\be
\lb{eitk}
\d (\t^{i}\{\e_{i}, \t^{(k)}\}) =  (k+1)(\a_{i}\t^{i-1}\e_{i}\t^{k} +
 \t^{i}\{\e_{i}, \t^{(k-1)}\}  )\;\;,\;\; k=1 \dots p-1 \;.
 \ee
We emphasize that, in general, these rules do not require any
$\e_{i}$ to be zero.

Thus the resulting infinitesimal paraconformal transformations
of the first order in $\lambda$ look as follows
\begin{eqnarray}
\delta z & = & \lambda(\o_{0} (z) -\frac{1}{\a_{p}}
\frac{1}{(\a_{p})!} \e_{0}(z) \t^{p}) \; ,
\nonumber  \\
\delta \t & = & \lambda(\e_{0}(z) + \frac{1}{p+1} \o_{0}'(z) \t +
\t^{2} \e_{2} + \dots +\t^{p} \e_{p}) \; ,
\lb{c23} \\
\delta (\t^{k}) & = & \lambda (\{\e_{0}, \t^{(k-1)}\} +
\frac{k}{p+1} \o_{0}'\t^{k} +
\t^{2}\{\e_{2}, \t^{(k-1)}\} + \dots +
\t^{p+1-k}\{\e_{p+1-k}, \t^{(k-1)}\}\;)\;.
\nonumber
\end{eqnarray}

To obtain generators of the transformations (\ref{c23}),
it is convenient to define new operators ${\cal J}_{0}$ and $\bd$
 acting on $\t^{k}$ in the following way
\begin{eqnarray}
{\cal J}_{0}(\t^{k}) & = & k \t^{k}\;,\lb{I0} \\
\e_{i}\bd (\t^{k}) & = &\{\e_{i}, \t^{(k-1)}\} \;, \lb{bd} .
\end{eqnarray}
The first operator is a generator of the automorphism group
of paragrassmann algebra [8]. The second one can be
interpreted as differentiation in certain other version ($\bar{N}$),
which is, in this sense, an `associate' to the original version ($N$).
So, in addition to (\ref{bd}), we assume that there exist a set of
non-zero numbers $\bar{\a}_{k}$ such that
\be
\lb{barak}
\bar{\d}(\t^{k}) = \bar{\a}_{k} \t^{k-1} \; .
\ee
This assumption is not too strong but, together with (\ref{bd}), it
leads to certain non-trivial restrictions on commutation of $\e_{i}$
and $\t$. We will not try to specify the associate version in general.
For the moment it is sufficient to know that for Version-$(1)_{q}$ and
Version-$(2)$ the associate versions are $(1)_{q^{-1}}$ and
Version $(2)$ itself, respectively.

Now it is convenient to introduce the operator
\be
\lb{sup}
Q = \bd - \frac{1}{\a_{p}}  \frac{\kappa}{(\a_{p})!}\t^{p} \d_{z}\;,
\ee
and choose  $\kappa = -\a_{p}\frac{(\a_{p})!}{(\bar{\a}_{p})!} $,
so that $Q^{p+1} = \d_{z}$ ($Q$ is an analog of the supersymmetry
generator and might be called the para-supersymmetry generator).

By virtue of these operators we can establish
the generators of the transformations (\ref{c23}) in any version
(from now on we omit the zero index of $\o_{0}$)
\begin{eqnarray}
T(\o) & = & \o \d_{z} +\frac{1}{p+1} \o' {\cal J}_{0} \; ,
\nonumber \\
              \lb{TGHany}
\G(\e_{0}) & = &
\e_{0} (\bd_{\t} + \frac{\t^{p}}{(\bar{\a}_{p})!} \d_{z} ) =
\e_{0} Q\;, \\
\H_{j}(\e_{j+1}) & = & \t^{j+1} \e_{j+1} \bd_{\t} \; .
\nonumber
\end{eqnarray}
These are the generators of the first generation $\{ {\cal L}^{(1)} \}$.
They must generate, through correct bilinear and multilinear brackets,
the entire algebra ${\cal CON}_{p}$
corresponding to the group ${\bf CON}_{p}$.

Let us now describe these generators in the preferred versions.\\
\it \\
Version-$(1)_{q}$.
\rm

Remember that this means  $\d \t = 1 + q\t \d$,   and therefore
 $\d \t^{k} = (k)_{q} \t^{k-1} + q^{k}\t^{k} \d$.  Then, assuming
%that all $\e_{i}$ are commutable with $\t$, i.e. having
%some numbers $r_{i}$ as the comfactors,
the commutation relations  $\e_{i}\t = r_{i}\t\e_{i}$,
where $r_{i}$ are some numbers,   we find from (\ref{eitk}) that all
these factors $r_{i}$ must be equal to $q$
(for $\e_{0}$ this can be seen from (\ref{c21}) since
$\a_{p} = (p)_{q} = - q^{p}$).
Now recalling the definition of $\bd$ (\ref{bd}) we find that
$$
\bd (\t^{k}) = (k)_{q^{-1}} \t^{k-1}\equiv q^{1-k}(k)_{q}\t^{k-1}\;,
$$
and so $\bd$ may be represented as $\bd = g^{-1}\d$,
where $g^{-1}$ is an automorphism of the paragrassmann algebra
$\Gamma_{\t}$ defined by  $g^{-1}(\t^{k}) = q^{-k}\t^{k}\;$.
It is easy to prove (see [8]) that
\begin{eqnarray}
g & = & q^{{\cal J}_{0}}  =  \d\t - \t\d \;\;,\nonumber   \cr
g^{-1} & = & q^{-{\cal J}_{0}}  =  \bd\t - \t\bd \;\;.\nonumber
\end{eqnarray}

Thus, we see that $\bd$ is in fact a differentiation in the associate
version $(1)_{q^{-1}}$. The generators $\G$ and $\H$ are represented as
\begin{eqnarray}
\G(\e) & = & \e ( \bd_{\t} + \frac{\t^{p}}{(p)_{q^{-1}}!} \d_{z} )
= \e Q\;\;,  \lb{kurd3} \\
\H_{j}(\e_{j+1}) & = & q^{-(j+1)}\e_{j+1}\t^{j+1} \bd_{\t} \; .
\end{eqnarray}
while $T$-generator has the same form as in Eqs. (\ref{TGHany}).

We would like to remind that the operator $\D$ (\ref{c2}) in
Version-$(1)_{q}$ and the para-supersymmetry generator $Q$ (\ref{kurd3})
have been introduced earlier in the context of the
%comparing with the paper [9], we exchange the operators $Q$ and $\D$.
fractional supersymmetry [9].                                \\
\it      \\
Version-\rm (2).

The unique property of this version is that the $\theta$-derivative
acts exactly like the standard one except the terms proportional to
$\t^{p}$
in the Leibniz rule and corresponding terms in the operator formulas.
This deviation from the standard rules is
not important in considering infinitesimal paraconformal
transformations (\ref{c23}).

The basic formula of Version-(2) required in this context is
\be
\lb{ver2a}
\d\t^{j}=j\t^{j-1}+\t^{j}\d -\frac{p+1}{(p+1-j)!} \t^{p}\d^{p+1-j}
\;,\;\;j=1 \dots p+1\;.
\ee
Applying it to the relations
\be
\lb{ver2b}
\d (\{\e_{0}, \t^{(k)}\}) =
 (k+1)\{\e_{0}, \t^{(k-1)}\} \;,\;\;k=1 \dots p-1\;,
 \ee
which are just (\ref{eitk}) for $\e_{0}$, one can easily get
by induction that
\be
\lb{ver2c}
\d(e_{0}\t^{k}) =k\;\e_{0}\t^{k-1}\;,\;\;k=1 \dots p-1\;.
\ee
 On the other hand, remembering the definition of $\bd$
(\ref{bd} -- \ref{barak}), we may regard (\ref{ver2b})
as the recurrence equation defining $\bar{\a}_{k}$
\be
\lb{ver2d}
\bar{\a}_{k+1}\;\d(\e_{0}\t^{k}) =
(k+1)\;\bar{\a}_{k}\;\e_{0}\t^{k-1}\;,\;\;k=1 \dots p-1\;.
\ee
On account of (\ref{ver2c}), we get $\bar{\a}_{k} =k$ for all
$k=1\dots p\;$ since $\bar{\a}_{1}=1$. Therefore $\bd \equiv \d$ and
the associate version is identical to Version-(2) itself.

This may be considered as a simplest example of a general procedure
for obtaining the associate version.
Note also that the relations (\ref{ver2c}) together with
$
\d(e_{0}\t^{p}) =-p^{2}\t^{p-1}\e_{0}\equiv
-\frac{p}{(p-1)!}\t^{p-1}\e_{0}\d^{p}\;(\t^{p})
$
can be summarized in a single operator formula
$$
\d\; \e_{0} = \e_{0} \d - \frac{p}{(p-1)!}\t^{p-1}\e_{0}\d^{p}
-\frac{1}{(p-1)!}\e_{0}\t^{p-1}\d^{p}\;.
$$
This is an example of how the commutation relations of an algebra with
many paragrassmann variables can look in a version other
than Version-(1).
In general, they look  monstrous and not suitable for computations.

The operator ${\cal J}_{0}$, standing in the $T$-generator, has
a very simple expression in Version-(2):
$
{\cal J}_{0} =\t\d\;,
$
and so the generators (\ref{TGHany}) look in this version in a quite
vector-like fashion
\begin{eqnarray}
T(\o) & = & \o \d_{z} +\frac{1}{p+1} \o' \t\d_{\t} \; ,
\nonumber \\
            \lb{TGH2}
\G(\e_{0}) & = & \e_{0} ( \d_{\t} + \frac{\t^{p}}{p!} \d_{z} )\;, \\
\H_{j}(\e_{j+1}) & = & \t^{j+1} \e_{j+1} \d_{\t} \; . \nonumber
\end{eqnarray}
Usually, to derive some identities between generators,
it is better to take them in Version-(2), in the form (\ref{TGH2}).
Version-(1) is better adapted to the computations
with many paragrassmann variables.    \\

\section{Paraconformal Algebras}
\setcounter{equation}0

Now let us turn to the algebra of the generators (\ref{TGHany}).
The commutators with $T$ are simple as they should be
due to the commutativity    of $\o$:
\begin{eqnarray}
\; [ T(\o) , \; T(\eta) ] & = & T(\o \eta' - \o' \eta ) \; ,
\lb{c25a} \\
\; [ T(\o) , \; \G(\e) ]   & = & \G( \o \e' - \frac{1}{p+1} \o' \e )
\; ,\lb{c25b} \\
\; [ T(\o) , \; \H_{j} (\xi ) ] & = & \H_{j} (\o \xi' +
\frac{j}{p+1} \o'\xi) \; .
\lb{c25c}
\end{eqnarray}
The rest of the commutators promise some subtleties. For instance,
the commutator of two $\G$-generators gives rise to a new generator
$\G^{(2)}$
\be
\lb{comGG}
[\G(\e),\;\G(\zeta)] \propto \G^{(2)}(\e\zeta -\zeta\e) + \frac{1}{2}
\H_{p-1}(\e\zeta'-\e'\zeta + \zeta'\e - \zeta\e')
\ee
This new generator has the form
\be
\lb{G2}
\G^{(2)}(\phi) = \phi Q^{2} +
\frac{1}{2}\phi'\frac{\t^{p}}{(\bar{\a}_{p})!}\bd
\ee
and contains a term $\phi \t^{p-1}\d_{z}$ , modifying $z$ on a quantity
proportional to $\t^{p-1}$, that is forbidden by the condition
(\ref{o123}). Note, however, that (\ref{o123}) is a condition on the
first-order variation, while $\G^{(2)}$ appears only in the second
order.  Similar effects occur when considering commutators of two
$\H$-generators or of $\G$ and $\H$, giving rise to the generators
$\H^{(2)}_{j}$. The latter have the form
$\H^{(2)}_{j}(\psi) \propto \psi \t^{j+1}\d^{2}$.

 $\G^{(2)}$ and  $\H^{(2)}_{j}$ are right those new generators of the
 second
generation, $\{{\cal X}^{2} \}$, expected to appear in the term of order
$\lambda^{2}$ in (\ref{genel}). They can be obtained directly
by repeating
the steps (\ref{DEk} -- \ref{TGHany}) but with accounting $\lambda^2$
terms in (\ref{c4}), (\ref{c4a}).

Similar procedure can be carried out (but the calculations
become more and more complicated) for
$\lambda^{M}\;,\;M=3\;,\;4\;,\;etc$. ,
giving rise to the generators of $M$-th generation
$ \{\G^{(M)} \;,\; \H^{(M)}_{j} \} \equiv \{ {\cal X}^{M} \} $.
Generators  $\G^{(M)}$
are right those that contain a term $\sim \t^{p+1-M}\d_{z}\;$,
which gives rise to non-zero $Z_{p+1-M}$ in (\ref{kurd1}).
Generators $\H^{(M)}_{j}$
are proportional to $\t^{j+1}\bd^{M}$ and do not affect $z$-coordinate.
They lead to a deviation of $\tilde{\t^{M}}$ from $(\tdt)^{M}$
on a quantity of order $\lambda^{M}$.
Probably, this could be interpreted as a shift of the version
during a paraconformal transformation.

New generators stop appearing at the order $\lambda^{p}$.
This fact could be explained by two circumstances.
First: the resource of possible combinations of $\t\;,\;\bd$ and
$\d_{z}$, which are the building blocks for generators, is exhausted
at the order $\lambda^{p}$. Second, (closely related to the first):
the algebra of the generators
$$
\{ {\cal L}^{p} \} =
\{ {\cal L}^{1} \} \cup
\{ {\cal X}^{2} \} \cup \dots \cup
\{ {\cal X}^{p} \} =
\{ T,\;\G,\;\H_{j}\;;\;\G^{2},\;
\H^{2}_{j}\;; \dots ;\;\G^{p} ,\;\H^{p}_{j} \}
$$
closes bilinearly.
So, if we denote by $A^{(M)}$ the linear shell of all the generators
of $M$-th generation, $\{ {\cal L}^{M} \} $, then the entire algebra
${\cal CON}_{p}$ can be represented as a $p$-filtered algebra, i.e.
the generators can be ordered as follows
$$
{\cal CON}_{p}
%\equiv A^{(\infty)}
 = A^{(p)} \supset A^{(p-1)} \supset \dots \supset
A_{(2)} \supset A^{(1)}  \; ; \;\;
[ A^{(M)}\;, \; A^{(K)} ]_{MK} \subset A^{(M+K)}\;.
$$
Each coset $A^{(M)}/A^{(M-1)} $ is based on
the generators ${\cal X}^{(M)}$.

Of course, this is a somewhat symbolic statement, because explicit
expressions for the bilinear brackets $[...]_{MK}$
are not known to us as yet, except the bracket $[...]_{11}$,
which is simply the commutator.
In general, it is not clear which combination of the para-generators
(this word applies to $\G$- and $\H$-type ones in any generation)
should be taken, because the true bracket should be determined by
some analog of the Hausdorff formula for the para-supergroup, and
the latter is not obtained so far.  Though taking the commutator of
two $\G$-generators seems to be more or less consistent with the
first order calculation, the commutator of, say, $\G$ and $\G^{(2)}$
seems hardly to have  any relation to the group ${\bf CON}_{p}$.
We would not like also to care about commutation properties of
$\e_{i}$, especially taking into account that for general
elements of the paragrassmann algebra of many variables the
commutation formulas are not specified. For $p=2$ we could do more as
one may see from the explicit formulas presented in Appendix. We will
give a detailed consideration of this simplest case in our next paper.

Just now we explore another approach, not restricted to special values
of $p$ and to particular versions of many-variable algebras.
Let us extract paragrassmann multipliers out
of all $\e$ and omit them, in analogy to what is done in
the supercase, leaving the arguments of all generators to be just
ordinary  functions commuting with everything. Then we have to
investigate the identities of such `deparagrassmannized'
(or, as we prefer to say, `bare') generators,
keeping in mind the hope that they will contain some hints about
the true structure of the true para-superalgebra and para-supergroup.
The complete and rigorous construction of these objects would
require a more    sharpened technique of handling
 paragrassmann algebras of many variables than we actually have.

So let us proceed with `bare' para-generators,
which will be denoted by the same letters but in more modest print.
Let us concentrate on $G$-generators,
for the reason that will be clarified below.

The identities generated by $G$'s can be described by three following
statements:

$1.\;\;\;$
The cyclic bracket of {\it any}  number of $G$-generators depends
only on the product of their arguments:
\be
\frac{1}{M}\{G(\e_{1}),\dots ,G(\e_{M})\}_{c} \equiv G^{(M)}(\eta)\;,
\;\;\eta = \e_{1}\cdot \dots \cdot \e_{M}
\lb{d5}
\ee
%This fact gives rise to a many multilinear constraints in the
%algebra $Con_{p}$.

$2.\;\;\;$
The similar statement is true for $G^{(M)}$:
\be
\frac{1}{K}\{G^{(M)}(\eta_{1}),\dots ,G^{(M)}(\eta_{K})\}_{c} =
G^{(KM)}(\zeta)\;,\;\;
\zeta = \eta_{1}\cdot \dots \cdot \eta_{K}
\lb{d6}
\ee

$3.\;\;\;$
\be
G^{(p+1)}(\o)\equiv T(\o)
\lb{d7}
\ee

These assertions can be proved by virtue of certain
identities in the algebra $\P$.
The third of them presents the simplest variant of the closure of the
algebra generated by $T$ and $G$ by the cyclic bracket, or the
{\it cyclator}, of $p+1\;\;G$-generators.

The generators $G^{(M)}$ must be treated as the `bare' variants
of $\G^{(M)}$. They have an elegant explicit form
\be
\lb{gm} G^{(M)}(\eta) = \eta\;Q^{M} + \frac{1}{M}\;\eta'
{\cal J}_{p+1-M}\;\; (M = 1, \dots , p+1) \;\;,
\ee
where ${\cal J}_{l}$ are certain generators of
paragrassmann algebra automorphisms acting as
$$
{\cal J}_{l}\left(\frac{\t^{k}}{(\bar{\a}_{k})!}\right) =
k\;\frac{\t^{k+l}}{(\bar{\a}_{k+l})!}\;.
$$

Another consequence of (\ref{d5} -- \ref{d7}) is
existence of a set of non-equivalent $(p+1)$-linear
brackets for composite $p+1$.  Really, if
$p+1 = \nu_{1}\cdot \dots \cdot \nu_{k}$ for some integer numbers
$\nu_{i}$, one can obviously replace $G^{(p+1)}$ in  (\ref{d7}) by the
$\nu_{1}$-linear cyclator of the generators
$G^{(\nu_{2}\cdots\nu_{k})}$.
Then, using (\ref{d6}), replace each of
$G^{(\nu_{2}\cdots\nu_{k})}$
by the $\nu_{2}$-linear cyclator of the generators
$G^{(\nu_{3}\cdots\nu_{k})}$,
and so on. As a result one gets a $(p+1)$-linear multi-cyclic
bracket of the generators $G$.
It is completely determined by the (ordered) sequence
$\nu = \langle \nu_{1},\dots ,\nu_{k}\rangle$
of the orders of sub-brackets (from outer to inner), and may
be labeled by the subscript $\nu$.
The subgroup of permutations that leaves
the bracket $\{\dots\}_{\nu}$ invariant will be denoted by $H_{\nu}$.
Its order, which coincides with the number of monomials in the
bracket, is
$$
N_{\nu}=\nu_{1}(\dots \nu_{k-1}(\nu_{k})^{\nu_{k-1}}\dots)^{\nu_{1}} \;.
$$
 It is curious to note that this number must be a divisor of $(p+1)!$.

Brackets corresponding to different sequences $\nu$ are linearly
independent. For example, for $p=5$ one can find
three `regular' brackets,
corresponding to the sequences $\langle 6 \rangle \;,\;
\langle 2,3 \rangle $ and $\langle 3,2 \rangle $,
which we represent symbolically as
$\{123456\},\{\{123\}\{456\}\}$ and $\{\{12\}\{34\}\{56\}\}$
(in this paragraph $\{ \dots \} = \{ \dots \}_{c} $).
They are evidently independent and containing 6, 18 and 24 terms
respectively. Less symmetrical brackets, like
$\{\{12\}\{34\}\{5678\}\}$ or like
$\{12\}3+\{23\}1+\{31\}2$, can be reduced to sums
of several classes of multi-cyclic brackets.

Thus the algebra generated by $T$ and $G$ can be established in
general form as
\begin{eqnarray}
\; [ T(\o) , \; T(\eta) ] & = & T(\o \eta' - \o' \eta ) \; ,
\nonumber \\
\; [ T(\o) , \; G(\e) ]   & = & G(\o \e' - \frac{1}{p+1} \o' \e )
  \;,\lb{Gb} \\
\{ G(\e_{0}), \; G(\e_{1}), \; \dots , G(\e_{p}) \}_{\nu} & = &
N_{\nu} T(\e_{0} \e_{1} \cdots \e_{p} ) \; .\nonumber
\end{eqnarray}
 Introducing the component generators
\be
\lb{c29}
L_{n} = T(z^{-n+1}) \; , \;\; G_{r} = G(z^{-r +1/(p+1)})
\ee
it can be written as
\begin{eqnarray}
\; [ L_{n} \; , \; L_{m} ] & = & (n-m) L_{n+m} \; , \nonumber \\
\; [L_{n} \; , \; G_{r} ] & = & (\frac{n}{p+1} -r) G_{n+r} \; ,
 \lb{c30} \\
\; \{ G_{r_{0}} , \dots , G_{r_{p}} \}_{\nu} & = & N_{\nu}
L_{\sum r_{j} } \; .\nonumber
\end{eqnarray}
A particular case of this algebra, when $\nu$ denotes the brackets
with all permutations of $G$'s and $N_{\nu}=\frac{(p+1)!}{p}$,
have been presented in the papers [7], [8].

It must be noted that the generators of the algebra (\ref{Gb})
possess a general representation depending on an arbitrary
`para-conformal' weight $\Delta$ (cf. (\ref{v11a})).
In the case of Version-(2), the generalization of the formulas
(\ref{TGH2}) looks as
\begin{eqnarray}
T(\o) & = & \o \d_{z} +\frac{1}{p+1} \o' (\t \d_{\t} + \Delta) \; ,
\nonumber \\
G(\e) & = & \e ( \d_{\t} + \frac{\t^{p}}{p!} \d_{z} )
 - \frac{\Delta}{p} \e' \frac{\t^{p}}{p!} \; . \lb{c24}
\end{eqnarray}

The algebra (\ref{Gb}), or (\ref{c30}), will be called below
{\em the paraconformal}, or $Con_{p}$, and its central extensions will
be presented in the next section.

Before turning to this task we have to make some comments on the
$H$-generators. Really, if the algebra $Con_{p}$ pretends to be a bare
form of ${\cal CON}_{p}$, it has to deal with all the
generators $H_{j}$ and $H^{(M)}_{j}$ , as well as with $T\;,\;G$ and
$G^{(M)}$. The problem is in the following. For generators $G$,
the structure of the identities (\ref{d5} -- \ref{d6}) and the form of
$G^{(M)}$ is practically fixed by the requirement, that the argument of
$G^{(M)}$ must be the product of the arguments of the correspondent
$G\;$'s. For $H$-generators,
the similar requirement is almost always fulfilled automatically,
and, therefore, it is not clear, which combination should be
considered as the right definition of  $H^{(M)}$.
Then, there exist $p-1$ generators  $H^{(M)}_{j}$
in each of $p$ generations $M$, so the number of different
identities blows up as $p$ increases. All this makes writing correct
relations with $H$-generators a rather subtle problem.

To illustrate the situation, consider the simplest
case $p=2$ with one generator $H(\xi)$.
The complete paraconformal algebra $Con_{2}$ may be written as
\be
\lb{con2h}
\begin{array}{rcl}
\; [ T(\o) , \; T(\phi) ] & = & T(\o \phi' - \o' \phi ) \; ,\cr
\; [ T(\o) , \; G(\e) ]   & = & G( \o \e' - \frac{1}{3} \o' \e )\;, \cr
\; [ T(\o) , \; H(\xi ) ] & = & H(\o \xi' +\frac{1}{3} \o'\xi) \;,\cr
\{G(\e), G(\zeta), G(\eta)\}_{c} & = & 3\; T(\e\zeta\eta)\;,\cr
\{G(\e), G(\zeta), H(\xi)\}_{c} & = & G(\e\zeta\xi)\;,\cr
\{G(\e), H(\sigma), H(\xi)\}_{c} & = & H(\e\sigma\xi)\;.
\end{array}
\ee
$G^{(2)}$-generator has the usual form (\ref{d5}).
$H^{(2)}$-generator can be defined by
\be
\lb{h21}
  G(\eta)H(\xi)+q^{1/2}H(\xi)G(\eta)=H^{(2)}(\eta \xi)\;,
\ee
so that
\be
\lb{h22}
G(\zeta)H^{(2)}(\tau)+q^{-1/2}H^{(2)}(\tau)G(\zeta)=G(\zeta \tau)\;.
\ee
Here $q$ denotes a primitive cubic root of unity,
but this has no connection to the $q$-version.
Note that there are no particular reasons for defining $H^{(2)}$ as
above, except conciseness of the formulas (\ref{h21}) and (\ref{h22}).

The cyclic brackets, similar to those in (\ref{con2h}),
also exist for $p>2$ but their number increases with $p$ very fast
due to the growing number of $H$-generators.
So the problem of a correct
description of the $H$-sector in the algebra $Con_{p}$
looks rather messy. It can hardly be solved
without using a Lie-type theory of para-supergroups.
For this reason we have excluded the $H$-generators
in our treatment of the algebras $Con_{p}$.
The other reason is that the $H$-generators are irrelevant to
constructing central extensions, as will be clarified in
the next section.

\section{Central Extension of the $Con_{p}$ Algebra}
\setcounter{equation}0

A geometric meaning of the algebra $Con_{p}$ becomes practically
obvious after noting that the arguments $\o$ ,
$\e$ of the generators (\ref{c24})
can be considered not as ordinary functions but as
$\lambda$-differentials of suitable weights.
The general rule is that generators representing the currents of
the spin $s$ (conformal dimension $s$) must have the
differentials of weight $\lambda = 1-s$ as their arguments.
So, for $T$ of dimension $2$ we have $\o \in \F^{-1}$ and for $G$ of
dimension $\frac{p+2}{p+1}$ we have $\e \in \F^{-\frac{1}{p+1}}$.
Here and below we denote by $\F^{\lambda}$
the space of the $\lambda$-differentials
$$
\F^{\lambda} = \{ \o(z) \; : \;\; \o(z) \mapsto (\tdz)^{\lambda}
\o(\tdz) \; ,
\;\; {\rm when} \;\; z \mapsto \tdz \} \; .
$$
The algebra of the generators is then determined by suitable
differential operators relating the differentials of different weights.
For the generators $T$ and $G$ we may write symbolically
\be
\begin{array}{rll}
\; [T(\o_{1}),T(\o_{2})] & = & T(l(\o_{1},\o_{2}))    \\
\; [T(\o)    ,G(\e)] & = & G(m(\o,\e))            \\
\; \{G(\e_{0}),\dots ,G(\e_{p})\}_{\nu} & =
& T(n(\e_{0},\dots ,\e_{p}))\;,
\end{array}
\lb{d1}
\ee
where the operators $l$, $m$, $n$ act as follows:
\be
\begin{array}{rcl}
l:\F^{-1}\Lambda \; \F^{-1} & \rightarrow & \F^{-1}  \\
	 (\o_{1}\;,\;\o_{2}) & \mapsto & \o_{1}\o^{\prime}_{2}-
\o^{\prime}_{1}\o_{2}  \\
	 l & = & d_{2}-d_{1}
\end{array}
\lb{d2}
\ee
%%%%
\be
\begin{array}{rcl}
m:\F^{-1}\Lambda \; \F^{-\frac{1}{p+1}} & \rightarrow &
\F^{-\frac{1}{p+ 1}}  \\
(\o\;,\;\e) & \mapsto & \o \e^{\prime}-\frac{1}{p+1}\o^{\prime}\e \\
	m & = & d_{2}-\frac{1}{p+1}d_{1}
\end{array}
\lb{d3}
\ee
%%%%
\be
\begin{array}{rll}
n:(\F^{-\frac{1}{p+1}}\times \dots \times \F^{-\frac{1}{p+1}})_{\nu}
& \rightarrow & \F^{-1}      \\
(\e_{0},\dots,\e_{p}) & \mapsto & \e_{0}\cdot \dots \cdot \e_{p} \;.
\end{array}
\lb{d4}
\ee
Here the symbol $d_{i}$ means differentiating the $i$-th multiplier;
the subscript $\nu$ reminds of the symmetry of the bracket.

Let us now turn to central extensions of $Con_{p}$. Being numbers,
central charges can arise from $\lambda$-differentials only as
residues of some 1-forms. Thus, all we have to do to get the central
extensions is to find out differential operators acting from the
left-hand sides of (\ref{d2}) and (\ref{d4}) to the sheaf
$\F^{1}/d\F^{0}$.
The first is the well-known and unique (modulo total derivative)
Gelfand-Fuks [19]
operator of differential order three: $\phi =d^{3}_{1}-d^{3}_{2}$.
The second must be of the order two and have the symmetry group
 of the  bracket, $H_{\nu}$. Thus for the simplest, cyclic bracket
we can construct $[\frac{p+1}{2} ]$ operators $\psi_{j}$:
\be
\psi_{j}=\sum_{i=1}^{p+1}d_{i}\:d_{i+j} \;\;,\;\;
j=1,\dots , \left[\frac{p+1}{2} \right]\;.
\lb{d8}
\ee
Not much more difficult is to construct the operators
$\psi^{(\nu)}_{j}$ corresponding to the bracket
$\{\dots \}_{\nu}$ with the symmetry $H_{\nu}$.
 For instance, the bracket of the type $\{\{123\}\{456\}\}$ admits
two operators
\begin{eqnarray}
\psi_{1} & = & d_{1}d_{2}+d_{2}d_{3}+d_{3}d_{1}+
         d_{4}d_{5}+d_{5}d_{6}+d_{6}d_{4} \;, \nonumber \\
\psi_{2} & = & d_{1}d_{4}+d_{2}d_{4}+d_{3}d_{4}+
	 d_{1}d_{5}+d_{2}d_{5}+d_{3}d_{5} \nonumber \\
	& + & d_{1}d_{6}+d_{2}d_{6}+d_{3}d_{6} \;.
\end{eqnarray}
And so on. The larger is the symmetry group of the bracket, $H_{\nu}$,
the smaller is the number of admissible central charges $E_{\nu}$.
For the regular brackets
$$
E_{\nu} = \sum_{k} \left[\frac{\nu_{k}}{2}\right]\;.
$$

The resulting extended algebra is
\be
\begin{array}{rll}
\; [T(\o_{1}),T(\o_{2})] & = & T(\o_{1}\o^{\prime}_{2}-\o^{\prime}_{1}
\o_{2})+C\phi(\o_{1},\o_{2})   \\
\; [T(\o)    ,G(\e)] & = & G(\o\e^{\prime}-
\frac{1}{p+1}\o^{\prime}\e) \\
\; \{G(\e_{0}),\dots ,G(\e_{p})\}_{\nu} & = & N_{\nu}T(\e_{0}\cdot
\dots \cdot \e_{p}) +
\sum_{j=1}^{E_{\nu}} c_{j}\psi^{(\nu)}_{j}(\e_{0},\dots ,\e_{p}) \;,
\end{array}
\lb{d9}
\ee
where $N_{\nu}$ is the number of the terms in the bracket. The central
charge $C$ can be expressed in terms of $c_{j}$ by
commuting the third line of Eq.~(\ref{d9}) with some $T(\eta)$ and
then comparing both sides of the resulting identity. Let us apply this
procedure to the cyclic bracket. Writing the generators in components
we get the algebra announced in the Introduction as $Vir_{p}$ :
\be \begin{array}{rll}
\;[L_{n},L_{m}] & = & (n-m)L_{n+m}+\frac{2}{p+1}(\sum_{j}c_{j})
		      (n^{3}-n)\delta_{n+m,0}   \\
\;[L_{n},G_{r}] & = & (\frac{n}{p+1}-r)G_{n+r}   \\
\;\{G_{r_{0}},\dots ,G_{r_{p}}\}_{c} & = & (p+1)L_{\Sigma r}-
\sum_{j}c_{j}(\sum_{i}r_{i}r_{i+j}+\frac{1}{p+1})\delta_{\Sigma r,0}
\end{array}
j=1,\dots , \left[\frac{p+1}{2}\right]
\lb{d10}
\ee

Note that the symmetry of the extension operators may be taken
wider than that of the bracket. That would be equivalent to
constraining some of the charges $c_{j}$. For example, there
exists a unique totally symmetric operator $\Psi =\sum_{i<j}d_{i}d_{j}$
that can be used with all kinds of brackets.
Unfortunately, this simple extension seems to be unsuitable for
constructing a non-trivial analog of the Verma module.

\section{Discussion and Conclusion}

Let us summarize the results and problems beginning with the results.

1. We have introduced the version-covariant fractional derivative, the
inverse operator (integral), and the Taylor expansion. This demonstrates
that the fractional derivative has many properties in common with the
superderivative and so may be indeed regarded as its generalization.

2. Transformations of the para-superplane preserving the form
of the fractional derivative $\D$ obey the transitivity condition
and form a group ${\bf CON}_{p}$ that is called the paraconformal group.
Simplest examples of the global paraconformal transformations are given
in Appendix.

3. The corresponding infinitesimal object, a `true' paraconformal algebra
${\cal CON}_{p}$, is related to the space of $p$-jets rather than to the
tangent space. ${\cal CON}_{p}$ is a $p$-filtered algebra
with generators in $p$ generations. Generators of $M$-th generation
do not occur in the order of smallness less than $M$.

The generators of the first generation are: the usual conformal
generator $T$ with the conformal weight $2$, the paraconformal generator
$G$ with the weight $\frac{p+2}{p+1}$, and the paragrassmann generators
$\H_{j} \; (j=1 \dots p-1 )$ with the weights $\frac{p+1-j}{p+1}$.
The generators of the $M$-th generation are: $\G^{(M)}$ with the weight
$1+\frac{M}{p+1}$ and $\H^{(M)}_{j}$ with the weight $\frac{p+M-j}{p+1}$.
The $\H$-type generators do not affect $z$-coordinate of the
para-superplane $(z,\t)$ but they are required by self-consistency of
the algebra.

4.  Algebra ${\cal CON}_{p}$  can be considered as a paragrassmann shell
of a `bare' (`skeleton') algebra  $Con_{p}$  also called paraconformal.
Its generators $T$, $G^{(M)}$, $H^{(M)}_{j}$ have as their arguments
ordinary functions (in fact, $\lambda$-differentials).  The connection
between ${\cal CON}_{p}$ and $Con_{p}$ is trivial
in the supercase ($p=1$) but it is not so clear for $p>1$.
We have systematically derived identities
in the algebra $Con_{p}$ that must encode some information about the
structure of the algebra ${\cal CON}_{p}$ but understanding of exact
relations between these two algebras is still lacking.

5. The algebra $Con_{p}$  can be closed in terms of $T$- and
$G$-generators  only (unlike ${\cal CON}_{p}$).
 There exist many multilinear identities with $G$-generators based on
the cyclic brackets of arbitrary order.
For composite $p+1$, they give rise to a set of non-equivalent
$(p+1)$-linear  brackets of $G$ closing to $T$.
This, by the way, makes evidence that the algebra $Con_{p}$ contains as
a subalgebra $Con_{r}$ when $(r+1)$ is a divisor of $(p+1)$ (in view of
drastic simplifications of the paragrassmann calculus and of the
paraconformal transformations occurring for prime integer $p+1$,
a detailed further treatment of this case is an obvious priority).

6. To each of these brackets there corresponds a set of basic central
extension  operators having the same symmetry as the bracket. The wider
is the symmetry, the smaller is the number of extensions. Constraining
the coefficients (the central charges), one can enlarge the symmetry of
the central term as compared to the bracket.

Here emerges a branching point for future development.

One way is to consider different brackets (and the identities of smaller
order as well) just as the identities in the same algebra $Con_{p}$,
and then to deal with the unique central extension that suits
all of them.
This corresponds to the extension generated by the totally symmetric
operator $\Psi$.

The other way is to forget all preliminaries about the infinitesimal
generators and paragrassmann algebras and to consider the algebras of
$T$ and $G$ with different brackets as independent infinite-dimensional
algebras, each having its own central extension. This approach might
appear fruitful
for simple brackets, like the cyclic ones. A right way is probably
somewhere in between and hence the problem of a lucky choice of
the symmetry breaking arises.

We think that `the right way' is that leads to a nontrivial Verma
module. In fact, a natural program to develop the theory is to define a
suitable analog of a Verma module over the algebra $Vir_{p}$ and
to search for degenerate
modules, Kac determinant and rational models. Unfortunately even first
steps appear to be nontrivial. Let us illustrate the problem by a
simple example.

Consider the algebra $Vir_{2}$ with the third line being taken with
a totally symmetric bracket:
$$
\{G_{r},G_{s},G_{t}\}_{sym} =
6L_{r+s+t}+C(r^{2}+s^{2}+t^{2})\delta_{r+s+t,0}.
$$
Assume that the constraint $F_{r+s} =
\{G_{r},G_{s}\}$ of the algebra $Con_{2}$
is preserved. Then for any acceptable positive $k$ we can write
two strings
$$
F_{\frac{2k}{3}}G_{-\frac{2k}{3}}+2F_{-\frac{k}{3}}G_{\frac{k}{3}}=
			6L_{0}+\frac{2}{3}C(k^{2}-1) ,
$$
$$
2F_{\frac{2k}{3}}G_{-\frac{2k}{3}}+F_{-\frac{4k}{3}}G_{\frac{4k}{3}}=
			6L_{0}+\frac{2}{3}C(4k^{2}-1) .
$$
Now defining a vacuum so that
$$
G_{>0}\approx 0\;\;,\;\;\;L_{0}\approx \Delta
$$
(we write $X\approx Y$ for $X\vert vac\rangle =Y\vert vac \rangle $ )
we immediately get a contradiction,
$$
F_{\frac{2k}{3}}G_{-\frac{2k}{3}}\approx
			6\Delta+\frac{2}{3}C(k^{2}-1) \approx
			\frac{1}{2}(6\Delta+\frac{2}{3}C(4k^{2}-1)) ,
$$
unless both $\Delta$ and $C$ are zero.
This is an evidence of a rather general phenomenon. Namely,
preserving constraints while keeping to a naive definition of the vacuum
is, as a rule, inconsistent with a nontrivial central charge
(and often with a nontrivial highest weight, as in the example). To
bypass this disaster one might either try to select an appropriate set
of the constraints to be preserved or to redefine the vacuum in a more
skillful way. Our attempts in this direction have
not produced anything valuable so far.

One might also suspect the $H$-generators might play a role
in defining the vacuum and the module.
However, the above example gives little support to this suspicion.

Some light on the topic might be thrown by investigating concrete
physical systems possessing paraconformal symmetry. But the algebraic
results of the present paper are hinting that quantizing such systems
will be rather ambiguous.

Thus, the current problems may be summarized in the following list:

1. The main theoretical problem is to find a rigorous connection
between the three constructed paraconformal objects: ${\bf CON}_{p}$,
${\cal CON}_{p}$ and $Con_{p}$. This problem requires further developing
the calculus for many paragrassmann variables.

2. It is not clear how to correctly include the $H$-generators into
the algebras $Con_{p}$ and $Vir_{p}$.

3. The main practical question is to find a non-trivial
Verma module over $Vir_{p}$.

Ending, we would like to note that paraconformal algebras are not of
a pure aesthetic interest. For example, on a Riemann surface of genus
$g$ the natural scale of conformal dimension is $\frac{1}{2(g-1)}$ ,
as a consequence of the Gauss-Bonnet theorem, and thus the natural
fractional derivative is of the same order, and the natural conformal
algebra would be $Vir_{2g-3}$ rather than $Vir_{1}=RNS$. One may also
speculate that using paraconformal algebras might drastically change
the critical dimensions in the string theory.

\section*{Appendix}

Here we present paraconformal transformations generalizing
translations, inversions and some other superconformal transformations.
To simplify formulas we consider the case $p=2$.

Let us try to write para-translations assuming $\tdt = \t + \e_0$, where
$\e_0$ is a paragrassmann number (independent of $z$ and satisfying
$\e_0^3 = 0$). The functions $Z_i$ can be found from Eqs.(\ref{kurd1}).
By applying the relation $[\tdz , \tdt] = 0$ and Eq.(\ref{Dtdtz})
we can find the commutation relation (\ref{c21}) and
$$\a_2(\a_2 - 1)\t {\e}_0^2 = \d({\e}_0^2 \t^2)$$
that immediately give
$$\a_2^2 - \a_2 + 1 = 0 .$$
This shows that either $\a_2 = -q$ or $\a_2 = -q^2$,
where $q$ is the prime
cubic root of unity. With $\a_2 = -q^2 = 1+q$ and choosing $k=1$ in
Eq.(\ref{c2}), the translations can be written in the form
$$\tdz = z+z_0 - q(\t \e_0^2 + \t^2 \e_0 )\; , \; \tdt =\t + \e_0 \; .$$
If we further assume that $\e_0 \t = q \t \e_0$ thus automatically
satisfying Eq.(\ref{c21}), we can find the operator generating
translations ($\bar{\a}_2 = -q = 1 + \bar{q}$)
$$\G(z_0 , \e_0)=e^{z_0 \d_z} (1+ \e_0 Q +{\bar{\a}_2}^{-1}
\e_0^2 Q^2)\; ,$$
$$\G(z_0 , \e_0) F(\t , z) = F(\tdt , \tdz)\;,$$
where $Q$ is defined by Eq.(\ref{sup}).
One can check that all conditions defining paraconformal transformations
are satisfied for the translations.

We may now apply these results to the integrals in Eqs.(\ref{tay1}),
(\ref{tay2}).
Assuming $\e_0 \t_i = q \t_i \e_0$ where $i=1, 2$ it is easy
to show that $I_2$ and $I_3$ are invariant under translations
(invariance of $I_1$ is obvious). However,
this does not give a naive generalization of  the properties
of the Taylor expansion in the Grassmann case. One can show
that the integrals $I_n$ can be simply expressed in terms of
$I_1$, $I_2$,
${\bar{I}}_3 =I_3 - (z_1 - z_2)/k$, of powers of $(z_1 - z_2)/k$, and of
one additional non-invariant expression $\t_1^2 \t_2^2$.
The transformation
properties of the coefficients in the Taylor expansion
(\ref{tay1}) are thus
not as obvious as in the Grassmann case. Possibly, our definition of the
translations is not quite suitable for this expansions.
In this connection,
we note that other definitions of translations are possible
but they require
specifying many-variable paragrassmann calculus which we consistently
avoid in this paper.

Finally, let us write more general paraconformal transformations.
With the above assumptions of $q$-commutativity between $\t$ and the
paragrassmann parameters, it is not difficult to show that the following
transformation satisfies all necessary conditions:
$$\tdt = \e_1 + (z/z_2)^{\l} \t \; ; $$
$$\tdz = [z_1 + z_2 (3\l +1)^{-1} (z/z_2)^{3\l + 1}] -
q(z/z_2)^{\l} \t \e_1^2 - q(z/z_2)^{2\l} \t^2 \e_1 \;.$$
Here $\l$, $z_1$, $z_2$ are complex parameters, $\e_1$ - a paragrassmann
number that $q$-commutes with $\t$, like $\e_0$. For $\l = 0$ we return
to the above translations. Applying translations to the general
transformations depending on arbitrary $\l$, we obtain rather general
paraconformal transformations depending on four complex numbers
$\l$, $z_0$,
$z_1$, $z_2$ and on two paragrassmann parameters $\e_0$ and $\e_1$. With
$\l = -2/3$ one can then derive a natural para-extension of
the $SL(2 , C)$
transformation. The above transformations may be further
generalized if we replace the power function
in the square brackets by an arbitrary function
of $z$ and express other powers of $z$ in terms of its derivative.
We think that these explicit formulas show that the para-superplane is
a rich but not hopelessly complicated thing worth of further study.

\section*{Acknowledgments}
This paper has been completed while one of the authors (A.T.F.) visited
Yukawa Institute of Theoretical Physics. A.T.Filippov wishes to express
his deep gratitude to Y.Nagaoka and T.Inami for a very kind hospitality
and support.
Useful comments by A.LeClair, S.Novikov, V.Spiridonov and S.Pakuliak are
also acknowledged.


\begin{thebibliography}{99}

\bibitem{1} A.B.Zamolodchikov, \it Rev. Math. Phys., {\bf 1}
\rm (1990)197; \\
L.Alvarez-Gaume, C.Gomez and G.Sierra, in: \it Physics and Mathematics
of Strings, \rm Eds. L.Brink, D.Friedan and A.Polyakov, World Sci.,
Singapore, 1990.

\bibitem{2} V.G.Knizhnik, \it Theor. Math. Phys.,
{\bf 66} \rm (1986)68; \\
M.Bershadsky, \it Phys. Lett., {\bf 174B} \rm (1986)285; \\
A.Schwimmer and N.Seiberg, \it Phys. Lett. {\bf 184B} \rm (1987)191; \\
E.Ivanov, S.Krivonos and V.Leviant, \it Nucl. Phys. {\bf B304} \rm
(1988)601.

\bibitem{3} V.A.Fateev and A.B.Zamolodchikov,
\it Sov. Phys. JETP {\bf 62} \rm (1985)215; \\
\it Theor. Math. Phys., {\bf 71} \rm (1987)451.

\bibitem{4} V.A.Rubakov and V.P.Spiridonov,
\it Mod. Phys. Lett. {\bf A3} \rm (1988)1337; \\
V.P.Spiridonov, \it J. Phys. {\bf A24} \rm (1991) L529.

\bibitem{5} S.Durand, R.Floreanini, M.Mayrand and L.Vinet,
\it Phys. Lett. {\bf 233B} \rm (1989)158.

\bibitem{6} Y.Ohnuki and S.Kamefuchi, \it Quantum Field Theory and
Parastatistics, \\
 \rm Univ. of Tokyo Press, 1982; \\
A.B.Govorkov, \it Sov. J. Part. Nucl. {\bf 14} \rm (1983)520.

\bibitem{7} S.Durand, \it Fractional Superspace Formulation
of Generalized Super-Virasoro Algebras,
\rm preprint McGill Univ. 92-30, 1992; HEP-TH 9205086.

\bibitem{8} A.T.Filippov, A.P.Isaev and A.B.Kurdikov,
 \it Mod. Phys. Lett. {\bf A7} \rm (1992)2129.

\bibitem{9} C.Ahn, D.Bernard and A.LeClair, \it Nucl. Phys.
{\bf B346} \rm (1990)409.

\bibitem{10} T.Nakanishi, \it Progr. Theor. Phys.
{\bf 82} \rm (1989)207.

\bibitem{11} D.Friedan, \it Notes on String Theory and Two Dimensional
Conformal Field  \\  Theory, \rm in:
\it Proceedings of the Workshop on Unified String Theories,\\
\rm Santa Barbara, 1985; \\
M.A.Baranov and A.S.Schwarz, \it JETP Lett. {\bf 42} \rm (1986)419.

\bibitem{11a} M.A.Baranov, I.V.Frolov and A.S.Schwarz, \it Theor. Math.
Phys. {\bf 79} \rm (1987) 64; \\
Ph.Nelson, \it Lectures on Supermanifolds and Strings, \rm in: \it
Particles, Strings and Supernovae, \rm Eds. A.Jevicki and C.I.Tan
W.Sci., Singapore, 1988; \\
C.Itzykson and J.-M.Drouffe, \it Statistical Field Theory,
\rm  Cambridge Univ. Press, Cambridge, 1989.

\bibitem{13} A.T.Filippov, A.P.Isaev and A.B.Kurdikov,
\it On Paragrassmann
Differential Calculus, \rm preprint JINR E5-92-392, Dubna, 1992; \\
\it Paragrassmann Differential Calculus, \rm HEP-TH 9210075, preprint
YITP/K-995, 1992, to be published in {\it Theor. Math. Phys.}.

\bibitem{14} S.Durand, R.Floreanini, M.Mayrand, V.Spiridonov and L.Vinet,
\it Parasupersymmetries and Fractional Supersymmetries, \rm in: \it
Proceedings of the XVIII-th International Colloquium on Group Theoretical
Methods in Physics,  \\ \rm Moscow, 1990.

\bibitem{15} N.Jacobson, \it Lie Algebras, \rm Interscience Publ.,
N.-Y.-London, 1962.

\bibitem{16} A.MacFarlane, \it J. Phys. {\bf A22} \rm (1989)4581; \\
L. Biedenharn, \it ibid. {\bf A22} \rm (1989)L873.

\bibitem{17} P.Brocker and L.Lander, \it Differential Germs and
Catastrophes,  \\ \rm Cambridge Univ. Press, Cambridge, 1975.

\bibitem{18} V.V.Vishnevskii, A.P.Shirokov and V.V.Shurygin, \it Spaces
over Algebras, \\ \rm Kazan University Publ., Kazan, 1985 (in Russian).

\bibitem{19} I.M.Gelfand and D.B.Fuks, \it Func. Anal. Appl. {\bf 3}
\rm (1969) 194; {\bf} 4 \rm (1970) 110.

\end{thebibliography}
\end{document}